\documentclass[twocolumn,superscriptaddress,floatfix,preprintnumbers, nofootinbib,hyperref]{revtex4} 
\pdfoutput=1
\usepackage[colorlinks=true,breaklinks=true]{hyperref}
\usepackage[normalem]{ulem}
\usepackage{amsmath, amsthm, amssymb, amsfonts,mathtools}
\usepackage{halloweenmath}
\makeatletter
\newcommand{\fmarki}{$\mathrightbat$}
\newcommand{\fmarkii}{$\xrightswishingghost{}$}
\newcommand{\fmarkiii}{$\mathcloud$}
\newcommand{\fmarkiv}{$\mathleftghost$}
\newcommand{\fmarkv}{$\mathleftbat$}               
\def\@fnsymbol#1{{\ifcase#1\or \fmarki\or \fmarkii\or \fmarkiii\or \fmarkiv\or \fmarkv\else\@ctrerr\fi}}
\makeatother

\usepackage[utf8]{inputenc}
\hypersetup{allcolors=[rgb]{0.0 0.0 0.6},linkcolor=[rgb]{0.75 0.05 0.05}}
\usepackage{epsfig}  
\usepackage{graphicx}  
\usepackage{subfigure}
\usepackage{slashed}       
\usepackage{url}
\usepackage{color}
\usepackage{multirow}
\usepackage{comment}
\usepackage{tikz-feynman}
\usepackage{xcolor}

\hypersetup{allcolors=[rgb]{0.0 0.0 0.6},linkcolor=[rgb]{0.75 0.05 0.05}}

\DeclareMathOperator{\cm}{cm}

\DeclareMathOperator{\MeV}{MeV}

\DeclareMathOperator{\s}{s}

\DeclareMathOperator{\erg}{erg}

\DeclareMathOperator{\km}{km}

\DeclareMathOperator{\kpc}{kpc}
\DeclareMathOperator{\g}{g}

\newcommand{\beq}{\begin{equation}}
\newcommand{\eeq}{\end{equation}}

\allowdisplaybreaks

\bibpunct{[}{]}{,}{n}{}{,}
\definecolor{ForestGreen}{RGB}{34,139,34}

\begin{document}
\preprint{LAPTH-054/23, CA21106}
\title{Comprehensive constraints on heavy sterile neutrinos\\ from core-collapse supernovae}

\author{Pierluca~Carenza}\email{pierluca.carenza@fysik.su.se}
\affiliation{The Oskar Klein Centre, Department of Physics, Stockholm University, Stockholm 106 91, Sweden}

\author{Giuseppe Lucente}
\email{giuseppe.lucente@ba.infn.it}
\affiliation{Dipartimento Interateneo di Fisica  ``Michelangelo Merlin,'' Via Amendola 173, 70126 Bari, Italy}
\affiliation{Istituto Nazionale di Fisica Nucleare - Sezione di Bari, Via Orabona 4, 70126 Bari, Italy}%
\affiliation{Institut f\"{u}r Theoretische Physik, Universit\"{a}t Heidelberg, Philosophenweg 16, 69120, Heidelberg, Germany}%
\affiliation{Universit\"{a}t Heidelberg, Kirchhoff-Institut f\"{u}r Physik, Im Neuenheimer Feld 227, 69120 Heidelberg, Germany}

\author{Leonardo Mastrototaro}
\email{lmastrototaro@unisa.it}
\affiliation{Dipartimento di Fisica ``E.R. Caianiello'', Università degli Studi di Salerno, Via Giovanni Paolo II, 132 - 84084 Fisciano (SA), Italy}
\affiliation{INFN - Gruppo Collegato di Salerno, Via Giovanni Paolo II, 132 - 84084 Fisciano (SA), Italy.}

\author{Alessandro~Mirizzi}
\email{alessandro.mirizzi@ba.infn.it}
\affiliation{Dipartimento Interateneo di Fisica  ``Michelangelo Merlin,'' Via Amendola 173, 70126 Bari, Italy}
\affiliation{Istituto Nazionale di Fisica Nucleare - Sezione di Bari, Via Orabona 4, 70126 Bari, Italy}%

\author{Pasquale Dario Serpico}
\email{serpico@lapth.cnrs.fr}
\affiliation{LAPTh, CNRS, USMB, F-74000 Annecy, France.}%

\begin{abstract}
Sterile neutrinos with masses up to $\mathcal{O} (100)$~MeV can be copiously produced in a supernova (SN) core, through the 
 mixing with active neutrinos. In this regard the SN 1987A detection of neutrino events has been used to put constraints on active-sterile neutrino mixing, exploiting the well-known SN cooling argument. We refine the calculation of this limit including
 neutral current interactions with nucleons, that constitute the dominant channel for sterile neutrino production. We also include,
 for the first time, the charged current interactions between sterile neutrinos and muons, relevant for the production
 of sterile neutrinos mixed with muon neutrinos in the SN core.  Using the recent modified luminosity criterion, we extend the bounds to the case where
 sterile states are trapped in the stellar core.   
 Additionally, we study the decays of heavy sterile neutrinos, affecting the  SN explosion energy and possibly producing a gamma-ray signal. We also illustrate the complementarity of our new bounds with cosmological bounds and laboratory searches. 
\end{abstract}
\date{\today}

\smallskip

\maketitle

\section{Introduction}

In the last decade, new theoretical ideas to address dark matter and other fundamental questions predict a dark sector composed of feebly interacting particles (FIPs) with sub-GeV masses and  very feeble interactions with Standard 
Model (SM) particles~\cite{Lanfranchi:2020crw,Agrawal:2021dbo,Alekhin:2015byh,Antel:2023hkf}. The most common approach to describe the interaction of the dark sector with the SM is through some {\it portal}. At this regard, the minimal portals mixing new dark sector states with gauge-invariant  combinations of SM fields are: vector (dark photons), scalar (dark Higgs), fermion (heavy neutral leptons) and pseudo-scalar (axions)~\cite{Lanfranchi:2020crw}. These portals are subject of  intense experimental investigations  with interesting plans for the next years~\cite{Lanfranchi:2020crw,Agrawal:2021dbo,Alekhin:2015byh,Antel:2023hkf}.

In this context, core-collapse supernovae (SNe) are recognized as a powerful laboratory not only to probe  fundamental neutrino properties~\cite{Mirizzi:2015eza,Janka:2012wk,Janka:2017vcp}, but also  the emission of FIPs (see, e.g., Refs.~\cite{Raffelt:1990yz,Raffelt:1996wa,Alekhin:2015byh,Antel:2023hkf}).
Indeed, for typical core temperatures $T \simeq \mathcal{O}(30)$~MeV, FIPs 
with masses up to $\mathcal{O} (100)$~MeV~\cite{Antel:2023hkf} can be abundantly produced in a SN core. Notably, the physics case of axions and axion-like particles~\cite{Carenza:2019pxu,Carenza:2020cis,Fischer:2021jfm,Lucente:2020whw,Caputo:2021kcv,Caputo:2022mah,Diamond:2023scc,Lella:2022uwi,Lella:2023bfb},  dark photons~\cite{Chang:2016ntp,DeRocco:2019njg,Sung:2019xie}
and dark Higgs~\cite{Dev:2020eam}
has been widely studied. 

In this paper, we will focus on another class of FIPs, namely 
heavy neutral leptons and in particular a 
heavy sterile neutrino, $\nu_4$, mostly a flavour-sterile one ($\nu_s$) with a (generally small) mixing with active neutrinos $\nu_\alpha$, $\alpha=e,\,\mu,\,\tau$.
These states have been
often introduced to explain the origin of neutrino masses~\cite{Merle:2017dhf,Boyarsky:2018tvu,Abazajian:2012ys,Abazajian:2019ejt}.
We remark that although the sterile neutrino scale considered here is not heavy for particle physics standards, it is so if compared to the current bounds on the mass scale in the active neutrino sector,  i.e. $m_\nu \lesssim 1$ eV. Therefore, we will use the adjective {\it heavy} in this sense. 

It is certainly not surprising that heavy sterile neutrinos, with masses well above the keV range, might have a strong impact on the SN dynamics~\cite{Shi:1993ee,Nunokawa:1997ct,Abazajian:2001nj,Hidaka:2006sg,Hidaka:2007se,Fuller:2008erj,Raffelt:2011nc,Arguelles:2016uwb,Suliga:2019bsq,Warren:2014qza,Warren:2016slz,Mastrototaro:2019vug,Syvolap:2019dat,Rembiasz:2018lok,Ray:2023gtu}. These particles, once produced in the hot SN core, escape from the star subtracting energy form the star. This energy-loss channel~\cite{Dolgov:2000pj,Dolgov:2000jw} 
might have a sizable impact on the duration of the neutrino burst. Requiring compatibilty with the SN 1987A observation in Kamiokande-II (KII)~\cite{Kamiokande-II:1987idp,Hirata:1988ad} and Irvine-Michigan-Brookhaven (IMB)~\cite{Bionta:1987qt,IMB:1988suc}  experiments (see Refs.~\cite{Li:2023ulf,Fiorillo:2023frv} for recent reanalyses of the SN 1987A neutrino signal) excludes a portion of the sterile neutrino parameter space.

This constraint has been recently re-evaluated in the \emph{free-streaming} regime in Ref.~\cite{Mastrototaro:2019vug}, considering \emph{weakly-mixed}  sterile neutrinos that escape the SN without interacting with stellar matter.
However, recent developments in  SN simulations and new proposals to improve FIP constraints from SNe suggest that the heavy-sterile neutrino limits can be significantly strengthened. 
It is worth noting that recent works, as Ref.~\cite{Mastrototaro:2019vug}, have considered the scattering of active neutrinos as the dominant channel for sterile neutrino production, neglecting the neutral current interactions with nucleons. The latter channel was expected to be suppressed, due to the Fermi-blocking associated with nucleon degeneracy in the SN core. Nevertheless, in the few cases where nucleon scattering was considered, as in the seminal papers~\cite{Dolgov:2000pj,Dolgov:2000jw}, the corresponding bounds were stronger than the ones obtained, e.g., in Ref.~\cite{Mastrototaro:2019vug}. However, since the treatment of these processes in Refs.~\cite{Dolgov:2000pj,Dolgov:2000jw} was cursory, it seems to us important to revisit and clarify this issue. Additionally,  from recent SN simulations \cite{Bollig:2017lki,Fischer:2020vie} it emerges that a population of muons is present in the core and neutrinos interact with them through charged current interactions. These interactions are especially relevant in enhancing the production of sterile neutrinos mixed with muon neutrinos, allowing for an improvement of the previous bounds. 

Furthermore, it is possible to constrain the $\nu_4$ parameter space by studying the energy deposited inside a SN via the electromagnetic decays of sterile neutrinos. This is relevant for massive sterile neutrinos, where various decay channels are possible. For example, the decay $\nu_{4}\to \pi^{0}\nu_{\mu}$ would deposit at least $135$~MeV of energy inside the SN~\cite{Fuller:2008erj,Rembiasz:2018lok}. 
At this regard, it has been recently shown in Ref.~\cite{Caputo:2022mah} that in order not to exceed the explosion energy observed in low-energy SNe, strong constraints can be placed on energy deposition induced by FIP decays.
This argument has been applied to the heavy sterile neutrino 
case in Ref.~\cite{Chauhan:2023sci}.
Finally, the flux of daughter particles produced outside the SN, especially $e^+$ and $\gamma$, may lead to strong bounds (see Ref.~\cite{Oberauer:1993yr} for a seminal study on $\nu_4$)
similarly to the ones recently  discussed in Refs.~\cite{Calore:2020tjw,Calore:2021klc,Calore:2021lih,Lella:2022uwi,Muller:2023vjm,Muller:2023pip} for the case of heavy axion-like particles.

Given these motivations, we devote  this work to strengthen the  existing bounds on heavy sterile neutrinos from SNe exploring different aspects: 
\begin{itemize}
    \item  including neutral current interactions of $\nu_4$  with nucleons; 
\item 
including charged current interactions of $\nu_4$  with muons;
\item  characterizing the trapping regime of $\nu_4$ verified at large mixing angles, adopting the the so-called ``modified luminosity criterion'', (see Refs.~\cite{Chang:2016ntp,Lucente:2020whw,Caputo:2021rux}). This recently proposed recipe allows one to extend the SN energy-loss bounds
also to the regime where $\nu_4$ are strongly interacting with matter;
\item considering (non)radiative decays of heavy  neutrinos, we strengthen  the cooling 
including the constraint from excessive energy deposition, following the method proposed in Ref.~\cite{Caputo:2022mah,Chauhan:2023sci}, and from an observable gamma-ray signal. 
\end{itemize}

The plan for this paper is as follows. In Sec.~\ref{sec:production_SN} we recall the heavy neutrino production in SNe and summarize the relevant production and absorption processes. Then in Sec.~\ref{sec:constraints} we discuss the different arguments presented in the literature to constrain FIPs from SNe and we apply them to the case of heavy $\nu_4$.
In Sec.~\ref{sec:comparison} we combine all our bounds and compare them with the other laboratory and cosmological constraints in the same mass range.
We conclude in Sec.~\ref{sec:conclusions}.
 In App.~\ref{Sec:Implementation of the code} we discuss the details of the evaluation of the charged and neutral current interactions involving $\nu_4$.

\section{Sterile neutrino production}
\label{sec:production_SN}

We limit ourselves  to heavy sterile neutrinos with masses $10~\MeV\lesssim m_4\lesssim 600~\MeV$~\cite{Asaka:2005an,Asaka:2005pn}, to avoid any possible resonant production which usually happens in the sub-MeV range~\cite{Raffelt:2011nc,Suliga:2020vpz,Arguelles:2016uwb}.  
In this mass range since the mixing of a sterile neutrino with electron neutrino is very constrained
(see, e.g., Ref.~\cite{Alekhin:2015byh}), 
we assume that the sterile neutrino is mixed dominantly with one active neutrino $\nu_\alpha$, with $\alpha=\mu,\tau$, such as
\begin{equation}
\begin{split}
\nu_\alpha &= U_{\alpha 1} \,\nu_\ell + U_{\alpha 4} \,\nu_4 \,\ ,   \\ 
\nu_s &= -U_{\alpha 4} \,\nu_\ell + U_{s4}\, \nu_4 \,\ ,
\end{split}
\end{equation}
where $\nu_\ell$ and $\nu_4$ are a light and the heavy mass eigenstate, respectively, $U$ is the unitary mixing matrix, linking mass and flavour states, and the most interesting parameter space corresponds to 
$|U_{\alpha 4}|^2\ll 1$, i.e.  $\nu_\ell$ is mostly active and $\nu_4$ is mostly sterile.

\begin{table*}[t]
    \centering
    \begin{tabular}{|c|c|c|}
    \hline
    Process & $|U_{\alpha 4}|^{-2}|\mathcal{M}|^2$ \\
    \hline
$\nu_{\alpha}+\bar{\nu}_{\alpha}\leftrightarrow\bar{\nu}_{\alpha}+\nu_4$ &$64G_F^2(p_1\cdot p_3)(p_2\cdot p_4)$ \\
$\nu_{\alpha}+\nu_{\alpha}\leftrightarrow\nu_{\alpha}+\nu_4$ &$32G_F^2(p_1\cdot p_2)(p_3\cdot p_4)$\\
$\nu_{\beta}+\bar{\nu}_{\beta}\leftrightarrow\bar{\nu}_{\alpha}+\nu_4$ &$16G_F^2(p_1\cdot p_3)(p_2\cdot p_4)$\\
$\nu_{\alpha}+\bar{\nu}_{\beta}\leftrightarrow\bar{\nu}_{\beta}+\nu_4$ &$16G_F^2(p_1\cdot p_3)(p_2\cdot p_4)$ \\
$\nu_{\alpha}+\nu_{\beta}\leftrightarrow\nu_{\beta}+\nu_4$ &$16G_F^2(p_1\cdot p_2)(p_3\cdot p_4)$ \\
$e^++e^-\leftrightarrow\bar{\nu}_{\alpha}+\nu_4$ &$64G_F^2[\tilde{g}_L^2(p_1\cdot p_4)(p_2\cdot p_3)+g^2_R(p_1\cdot p_3)(p_2\cdot p_4)-\tilde{g}_Lg_Rm_e^2(p_3\cdot p_4)]$\\
$\nu_{\alpha}+e^-\leftrightarrow e^-+\nu_4$ &$64G_F^2[\tilde{g}_L^2(p_1\cdot p_2)(p_3\cdot p_4)+g^2_R(p_1\cdot p_3)(p_2\cdot p_4)-\tilde{g}_Lg_Rm_e^2(p_1\cdot p_4)]$\\
$\nu_{\alpha}+e^+\leftrightarrow e^++\nu_4$ &$64G_F^2[g_L^2(p_1\cdot p_3)(p_2\cdot p_4)+\tilde{g}^2_R(p_1\cdot p_2)(p_3\cdot p_4)-\tilde{g}_Lg_Rm_e^2(p_1\cdot p_4)]$\\
$\nu_\alpha+N\leftrightarrow N+\nu_4$ &$|\mathcal{M}|^2_{AA}+|\mathcal{M}|^2_{VA}+|\mathcal{M}|^2_{VV}$\\
$\mu^-+N\leftrightarrow N'+\nu_4$ &$|\mathcal{M}|^2_{AA}+|\mathcal{M}|^2_{VA}+|\mathcal{M}|^2_{VV}$\\
$\mu^-+\nu_e\leftrightarrow e^-+\nu_4$ &$64G_F^2(p_1\cdot p_2)(p_3\cdot p_4)$\\
    \hline
    \end{tabular}
\caption{Squared matrix elements for sterile neutrino scattering processes  (assuming mixing with the species $\alpha$, and $\beta\neq\alpha$), summed over initial and final states, where $\tilde{g}_L=-\frac{1}{2}+ \sin^2 \theta_W$, $g_R=\sin^2 \theta_W$~\cite{Mastrototaro:2021wzl}. The symmetry factor $S=1/2!$ is already included when two identical particles are present in the same state (second row). The particles involved in each reaction are enumerated as $1+2\leftrightarrow 3+4$. The last two processes are valid only in the case of mixing with the muon neutrino, $\alpha=\mu$. The terms $|\mathcal{M}|^2_{AA}$, $|\mathcal{M}|^2_{VA}$ and $|\mathcal{M}|^2_{VV}$ are reported in App.~\ref{Sec:Implementation of the code}. In the $\mu \, N\leftrightarrow N'\, \nu_4$ matrix element we neglected the terms of higher order in the nucleon momenta, thus working at leading order in a non-relativistic approximation. Here, $N$ and $N'$ represent the different nucleons involved in the interaction in the initial and final states. Finally, for all the processes we have considered their corresponding charged conjugate for the $\bar{\nu}_4$ production (the last two processes are irrelevant because of the absence of $\mu^{+}$ in the SN core) and absorption.} 
\label{tab:scatteringt}
\end{table*}

In the SN core, sterile neutrinos are produced via the processes listed in Tab.~\ref{tab:scatteringt}. We characterize these processes closely following Ref.~\cite{Mastrototaro:2019vug}. We have neglected the bremsstrahlung process $NN\rightarrow NN\nu\bar\nu$ since it is always sub-leading in the interesting parameter space. Indeed, as a production channel, the computed luminosity according to the rate of~\cite{Raffelt:1996wa} is inferior to the one associated to the other processes in Tab.~\ref{tab:scatteringt}.
As an absorption channel, it is suppressed compared to $N\nu\rightarrow N\nu$ for obvious phase-space reasons. 

The production rate of sterile neutrinos per unit volume and energy can be written as 
\begin{equation}
\begin{split}
    \frac{{\rm d}^{2}n_{_4}}{{\rm d}E_4 {\rm d}t}&=\int \frac{{\rm d}^{3}p_{1}}{(2\pi)^{3}2E_{1}}\frac{{\rm d}^{3}p_{2}}{(2\pi)^{3}2E_{2}}\frac{{\rm d}^{3}p_{3}}{(2\pi)^{3}2E_{3}}\frac{4\pi E_4p_4}{(2\pi)^{3}2E_4}\\
    &(2\pi)^{4}\delta^{4}(p_1+p_{2}-p_{3}-p_4)|\mathcal{M}|^{2}_{12\rightarrow34}f_1\,f_2\,(1-f_3)\,,
    \end{split}
\label{eq:dnde}
\end{equation}
where $E_4$ and $p_4$ are energy and momentum of the sterile neutrino, $f_i$ is the distribution function of $i$-th particle involved in the process, $|\mathcal{M}|^2_{12\leftrightarrow 34}$ is the sum of the
squared amplitudes for collisional processes $1+2\leftrightarrow3+4$ relevant for
the sterile production/absorption, reported in Tab.~\ref{tab:scatteringt}. Given recent SN simulations including muons~\cite{Bollig:2017lki,Fischer:2020vie}, here we consider for the first time reactions involving muons, also reported in Tab.~\ref{tab:scatteringt}. Moreover,  we also include the neutral current interaction between neutrinos and nuclei, which results to be one of the main channels for the production and absorption of the heavy state. Despite the fact that its possible relevance was already pointed out in Ref.~\cite{Raffelt:1992bs}, in most literature it has been neglected, similarly to the process $\nu \,e^-\rightarrow\nu\,e^-$, due to the large assumed fermion degeneracy and Pauli blocking effect. The abundance and degeneracy of nucleons in the SN core can be assessed by considering that the nuclear medium is described in a relativistic mean-field (RMF) picture~\cite{Hempel:2014ssa}, according to which the nucleon distribution function $f_N$ is given by~\cite{Hempel:2014ssa}
\begin{equation}
    f_N(p) = \frac{1}{1+\exp\left[\left(\sqrt{p^2 + {m_N^*}^2} - \mu^*_N\right)/T\right]}\,,
\end{equation}
where $m_N^*=m_N+\Sigma_S$ is the effective nucleon mass, with $\Sigma_S$ the so-called nuclear scalar self-energy, and and $\mu^*_N$ the effective or kinetic chemical potential, defined as~\cite{Hempel:2014ssa}
\begin{equation}
    \mu^*_N = \mu_N - \Sigma_V\,,
\end{equation}
where $\mu_N$ is the nucleon chemical potential including the nucleon rest mass and $\Sigma_V$ is the RMF vector self-energy. Thus, nucleons have Fermi-Dirac distribution functions equivalent to a non-interacting system with effective chemical potentials $\mu^*_N$ and particle masses $m_N^*$ and their degeneracy can be estimated by introducing the degeneracy parameter $\eta_N$ defined as
\begin{equation}
    \eta_N = \frac{\mu^*_N - m_N^*}{T}\,.
\end{equation}
On the other hand, leptons in the SN simulations are described by the usual Fermi-Dirac distributions
\begin{equation}
    f_l(p) = \frac{1}{1+\exp\left[\left(\sqrt{p^2 + m_l^2} - \mu_l\right)/T\right]}\,,
\end{equation}
with $m_l$ their bare mass and $\mu_l$ their chemical potential, leading to the degeneracy parameter
\begin{equation}
    \eta_l = \frac{\mu_l - m_l}{T}\,.
    \label{eq:etal}
\end{equation}
We mention here that electrons in the plasma acquire an effective mass that for typical SN conditions (${\mu_e \gg T \gg m_e}$) can be written as ${{m_e^*}^2 = e^2\,(\mu_e^2 + \pi^2\,T^2)/8\pi^2}$~\cite{1992ApJ...392...70B}. This expression leads to $m_e^*\lesssim \mathcal{O}(10)~\MeV \ll \mu_e$ in the SN core. Thus, using $m_e$ or $m_e^*$ marginally affects the evaluation of $\eta_e$ in Eq.~\eqref{eq:etal}~\cite{Lucente:2021hbp} and, consistently with our benchmark SN model described in the following, we neglect $m_e^*$ in our analysis.
Particles $i$ in the plasma are non-degenerate if $\eta_i < 0$, while they are fully degenerate for $\eta_i\gg 1$ and only partially degenerate for intermediate values of $\eta_i$~\cite{Raffelt:1996wa}.

We compute the sterile neutrino production using as a benchmark an 18~$M_{\odot}$ progenitor mass (roughly consistent with Sanduleak-69 202, the progenitor of the SN 1987A) obtained using a 1D spherically symmetric and general relativistic hydrodynamics model, based on the {\tt AGILE BOLTZTRAN} code~\cite{Mezzacappa:1993gn,Liebendoerfer:2002xn}, including muons. While we expect that these simulations capture the basic physics of the phenomenon, differences of a factor of a few can be associated to the implementation scheme of the neutrino microphysics, general relativistic effects, multi-dimensionality, etc. We think that this constitutes the dominating systematic error in the derived bounds.

\begin{figure}[t!]
    \centering
    \includegraphics[width=1.\columnwidth]{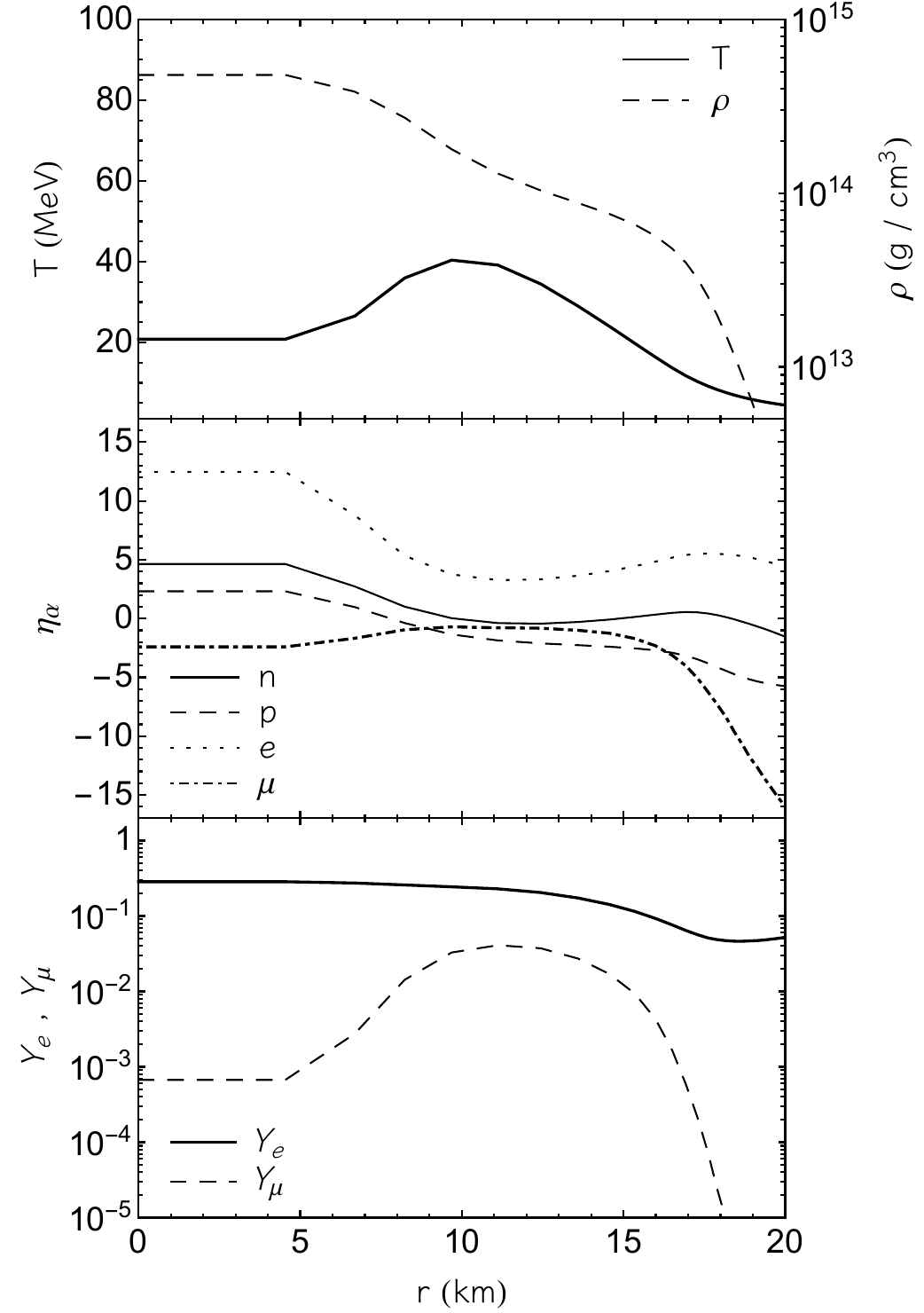}
    \caption{\emph{Upper panel}: Radial profiles of  the temperature $T$ (solid black line) and density $\rho$ (dashed line) in the SN core. \emph{Middle panel}: Radial profile for the degeneracy parameters of neutrons $\eta_n$ (solid black line), protons $\eta_p$ (dashed black), electrons $\eta_e$ (dotted black) and muons $\eta_\mu$ (dot-dashed red). \emph{Lower panel}: Radial profiles of the electron fraction $Y_e$ (solid line) and muon fraction $Y_\mu$ (dashed line). All panels refer to the post-bounce time $t_{\rm pb} = 1~$s.
    }
    \label{fig:T}
\end{figure}

We show in Fig.~\ref{fig:T} the thermodynamical conditions for our benchmark SN model in the inner core ($r\lesssim 20~\km$) at the post-bounce time $t_{\rm pb}=1~\s$. The upper panel shows the temperature $T$ (solid black line), with a peak $T \sim 40~\MeV$ at $r\sim10~\km$, and the matter density $\rho$ (dashed black), with a maximum $\rho \sim 4\times 10^{14}~\g~\cm^{-3}$ at the center and decreasing at larger radii. The central panel shows the fermion degeneracy parameters $\eta_\alpha$ for nucleons $\alpha=n,\,p$ and leptons $\alpha=e,\,\mu$. In the very inner core ($r\lesssim 5~\km$), neutrons (solid black line) are degenerate ($\eta_n \approx 5$) and protons (dashed black line) are partially degenerate ($\eta_p \approx 2$). For larger radii $(r\gtrsim 10~\km)$, the nucleon degeneracy decreases, implying non degenerate protons $(\eta_p < 0)$ and only partially degenerate neutrons $(\eta_n \lesssim 1)$. On the other hand, throughout the SN core, electrons (dotted black line) are highly degenerate ($\eta_e \gtrsim 5$) and muons (dot-dashed black line) are non-degenerate ($\eta_\mu < 0$). This implies that the $\nu\,e^- \to \nu e^-$ process is suppressed by the electron degeneracy, while neutral current interactions with nucleons cannot be neglected, at least in the outer layers of the core. Moreover, we checked that the conclusions concerning the $\eta_e$ parameter are unchanged (with a discrepancy lower than $5\%$) even considering the effective electron mass due to QED at finite temperature and density, yielding  $m^*_e\sim\mathcal{O}(10)~\mathrm{MeV}$ at $r\lesssim\mathcal{O}(10)~\mathrm{km}$~\cite{1992ApJ...392...70B,Lucente:2021hbp}. 
Finally, in the lower panel we present the electron $Y_e$ (solid line) and muon $Y_\mu$ (dashed line) abundance with respect to the nucleon one, with $Y_\alpha=n_\alpha/n_B$, where $n_\alpha$ is the density per unit volume for the particle $\alpha=e,\mu$ and $n_B$ is the baryon number density. We realize that the muon abundance around the peak of the temperature can be $\mathcal{O}(10\%)$ of the nucleon one. Therefore, for definiteness, we evaluate the sterile neutrino production by taking into account also processes involving muons.

We compute the production rate for sterile neutrinos by reducing the nine-dimensional integral in Eq.~\eqref{eq:dnde} to a three-dimensional one following the procedure in Ref.~\cite{Hannestad:1995rs}.  As an example, in App.~\ref{Sec:Implementation of the code}, we show how it is possible to write the interaction matrix elements for the charged current process process $\mu\,N\leftrightarrow N\, \nu_4$ and the neutral current interaction $ \nu_\alpha\, N \leftrightarrow N\,\nu_4 $ using the formalism in Ref.~\cite{Hannestad:1995rs}. The same procedure can be applied to evaluate the interaction matrix elements for the other processes we consider.

\section{SN Constraints on heavy sterile neutrinos}
\label{sec:constraints}
\subsection{Cooling bound}
From the observation of the ${\bar\nu}_e$ neutrino burst from SN~1987A~\cite{Kamiokande-II:1987idp,Hirata:1988ad,Bionta:1987qt,IMB:1988suc}, it is possible to  infer the temporal evolution of the  neutrino lightcurve. Despite the sparseness of the data, the duration of the neutrino burst extending over 10~s is in agreement with the expectations from the SN cooling via 
neutrinos~\cite{Raffelt:1996wa}. Therefore, from the  SN~1987A neutrino data there is no evidence of a dominant emission of FIPs, that would have significantly shortened the duration of the neutrino burst~\cite{Raffelt:1990yz,Raffelt:1996wa}.\,\footnote{ In Ref.~\cite{Fiorillo:2023frv} it has been noticed that the latest three events of SN~1987A observed at $t_{\rm pb}\sim 10$~s are in tension with the state-of-the-art SN simulations. Inferring the luminosity bound only on the events observed by IMB up to $t_{\rm pb}\sim 5$~s, one expects a relaxation of the bound by a factor two~\cite{Raffelt:1987yt}.}

\begin{figure}[t!]
    \centering
    \includegraphics[width=1.\columnwidth]{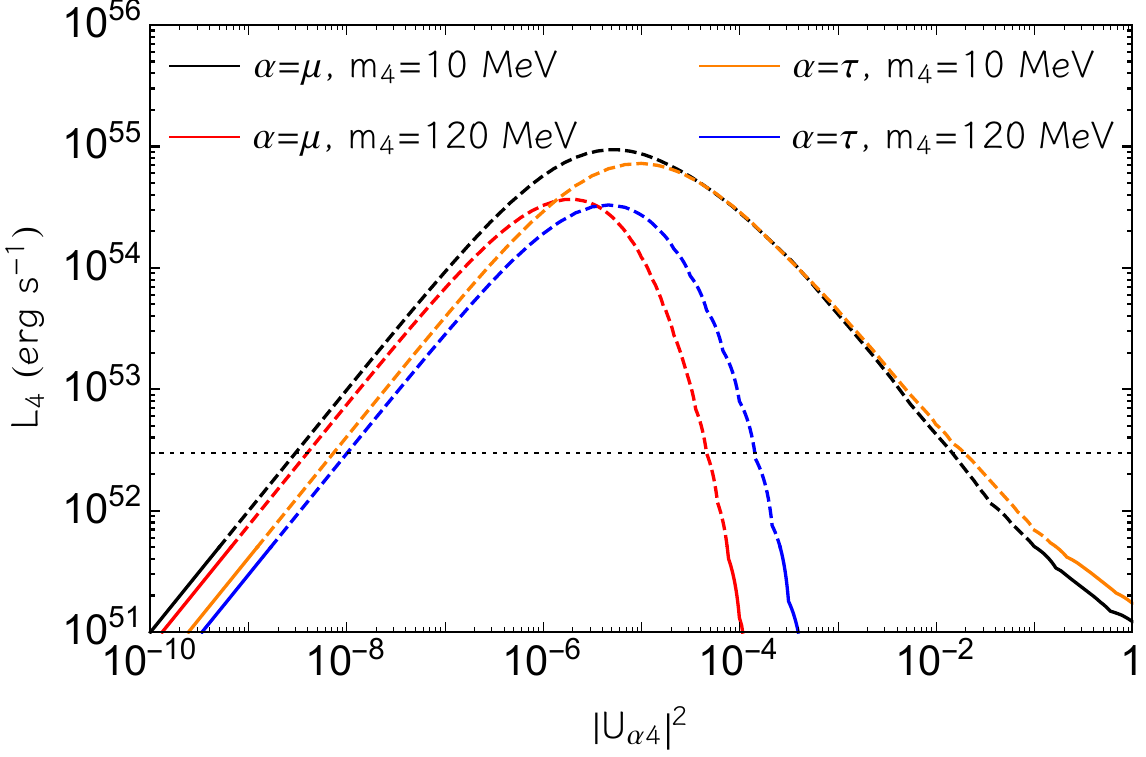}
    \caption{Sterile neutrino luminosity as a function of $|U_{\alpha 4}|^2$ for $\alpha=\mu$, $m_4=10~\rm MeV$ and $m_4=120~\rm MeV$ (black and red lines, respectively) and the same for $\alpha=\tau$ (orange and blue lines, respectively). The horizontal dotted line corresponds to the limit value of $L_\nu =3\times10^{52}~\erg~\s^{-1}$. Dashed lines are used for values of the mixing where the sterile neutrino luminosity exceeds the luminosity of the species $\nu_\alpha$ mixed with $\nu_4$. }
    \label{fig:L}
\end{figure}

In order to avoid a significant shortening of the observed neutrino burst  due to FIP emission, one should require that the luminosity of the exotic particles should be less then the one carried by neutrinos. Namely, for our fiducial model at $t_{\rm pb}=1$~s one has~\cite{Raffelt:1990yz,Raffelt:1996wa,Caputo:2021rux} 
\begin{equation}
     L_{\rm FIP}\lesssim L_\nu \equiv 3\times 10^{52}~\mathrm{{erg}\,\ {s}^{-1}} \,\ .
     \label{eq:constraintcooling}
\end{equation}
Our goal is to use this constraint to exclude values of the $\nu_{4}$ mixing with muon neutrinos ($|U_{\mu 4}|^2$) and tau neutrinos ($|U_{\tau 4}|^2$). We do not consider the case of mixing with the electron flavour, since in this case the parameter space is overconstrained.
We adopt the ``modified luminosity criterion''~\cite{Chang:2016ntp,Lucente:2020whw,Caputo:2021rux} to smoothly interpolate between the regimes in which sterile neutrinos are so weakly interacting that they freely escape from the SN
(i.e. weak-mixing regime with $|U_{\alpha 4}|^2\lesssim 10^{-5}$, see Fig.~\ref{fig:L}), also known as free-streaming regime, and a regime of stronger interactions with matter
(i.e. strong-mixing regime with $|U_{\alpha 4}|^2\gg 10^{-5}$, see Fig.~\ref{fig:L}), when they are trapped in analogy with active neutrinos.
In this formalism, the $\nu_{4}$ luminosity 
is~\cite{Chang:2016ntp,Lucente:2020whw,Caputo:2021rux,Caputo:2022rca} 
\begin{equation}
L_{4}= 4\pi \int_0^{\infty} {\rm d} r\, r^2\, \alpha^2 (r) \int {\rm d}E_4 E_4 \frac{{\rm d}^2{ n}_4}{{\rm d} E_4{\rm d}t} \langle e^{-\tau(E_4',r)}\rangle\,,
\label{eq:la}
\end{equation}
where $\alpha$ is the lapse factor to account for the gravitational redshift and the exponential suppression $e^{-\tau}$ takes into account the possibility of $\nu_{4}$ absorption inside the SN. In particular, $\langle\,e^{-\tau}\rangle$ is a directional average of the absorption factor~\cite{Caputo:2021rux,Caputo:2022rca,Lucente:2022vuo,Carenza:2023lci}
\begin{equation}
    \langle e^{-\tau(E',r)} \rangle = \frac{1}{2} \int_{-1}^{+1} {\rm d}\mu\, e^{-\int_0^{\infty} {\rm d}s \,\lambda^{-1} (E',\sqrt{r^2+s^2+2\,r\,s\,\mu})}\,,
    \label{eq:absfac}
\end{equation}
where $\lambda$ is the sterile neutrino mean-free path (mfp), $E'=E \,\alpha(r)/\alpha\left(\sqrt{r^2+s^2+2\,r\,s\,\mu}\right)$ is the $\nu_{4}$ redshifted energy, $\mu=\cos\beta$ and $\beta$ is the angle between the outward radial direction and a given ray of propagation along which $s$ is integrated.
We emphasize that, lacking self-consistent SN simulations including the feedback due to the emission of sterile neutrinos, as the rest of the literature (implicitly) does, we also resort to an extrapolation whenever the extra neutrino luminosity is comparable with or larger than the luminosity of the species $\nu_\alpha$ mixed with $\nu_4$ (see dashed lines in Fig.~\ref{fig:L}). The results at values much larger than the active neutrino luminosity are however only nominal, and not essential in obtaining the bound. Yet, it is conceivable that this limitation may introduce a factor of a few uncertainty in the limits from the cooling argument.

\begin{figure}[t!]
    \centering
    \includegraphics[width=1\columnwidth]{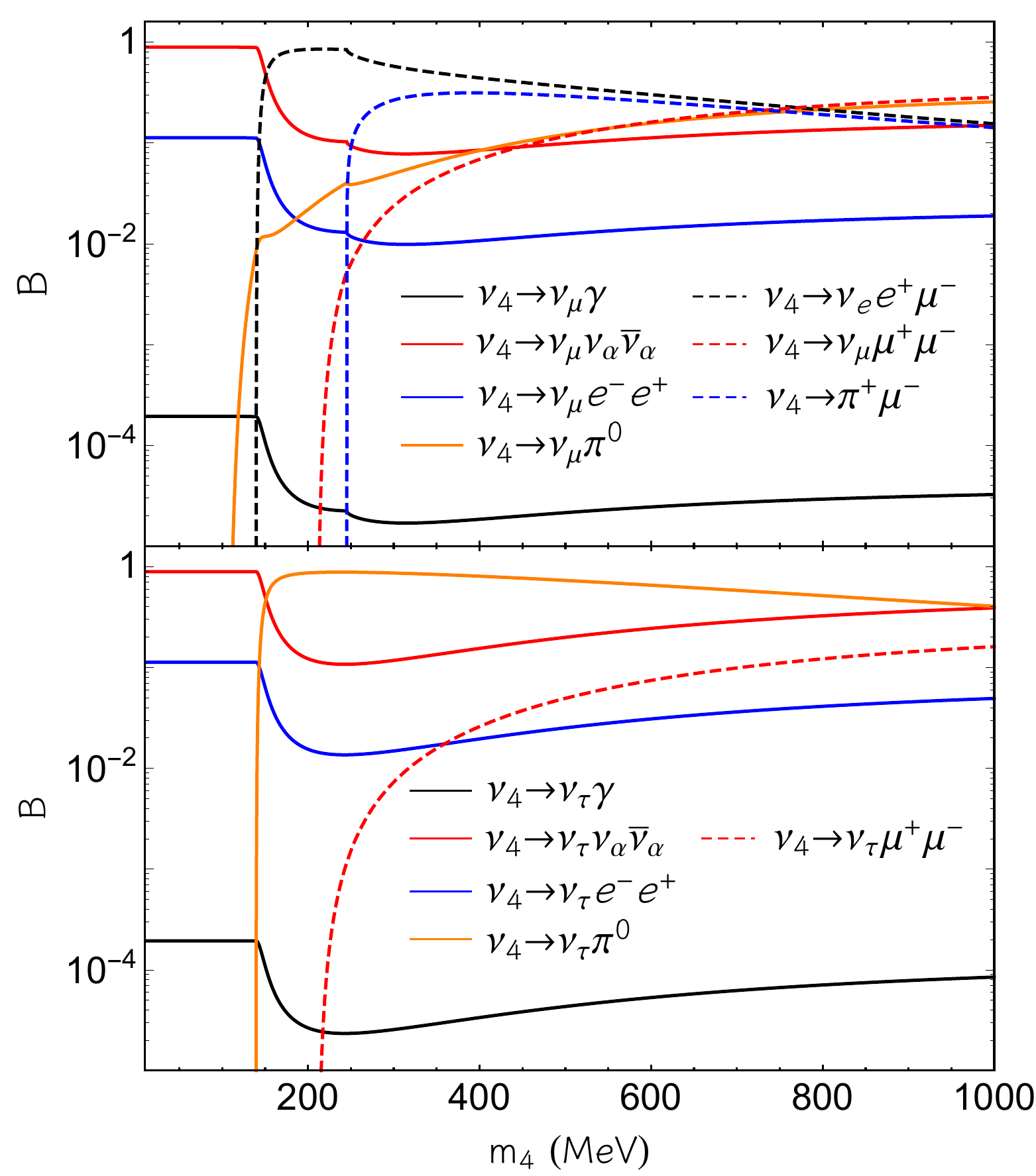}
    \caption{Branching ratios for the relevant decay channels listed in Tab.~\ref{decayt} as a function of the sterile neutrino mass $m_4$ for a mixing with $\nu_\mu$ (upper panel) and with $\nu_\tau$ (lower panel).}
    \label{fig:BR}
\end{figure}

\begin{table*}[t!]
    \centering
    \begin{tabular}{|c|c|c|}
    \hline
    Process & $\Gamma/G_F^2m_4^3|U_{\mu 4}|^2$ & Threshold (MeV)\\
    \hline
    $\nu_4\rightarrow\nu_\mu\gamma$& $9\alpha m_4^2/2048\pi^4$ & $0$\\
    $\nu_4\rightarrow\nu_\mu\nu_\mu\bar{\nu}_\mu$& $m_4^2/384\pi^3$ & $0$\\
    $\nu_4\rightarrow\nu_\mu\nu_{e(\tau)}\bar{\nu}_{e(\tau)}$& $m_4^2/768\pi^3$ & $0$\\
    $\nu_4\rightarrow\nu_\mu e^+e^-$& $(\tilde{g}_L^2+g_R^2)m_4^2/192\pi^3$ & $1.02$\\
    $\nu_4\rightarrow\nu_e e^{+} \mu^{-}$ & $
    \begin{multlined}\\
    m_4^2/384\pi^3\Big(
    2\left(1-m_\mu^2/m_4^2\right)\left(2+9m_\mu^2/m_4^2\right)\\+2m_\mu^2/m_4s^2 
     (1-m_\mu^2/m_4^2)\\ \left(-6-6m_\mu^2/m_4^2+m_\mu^4/m_4^4+6\log{m_\mu^2/m_4^2}\right)\Big)\\
    \end{multlined}
    $ & $106.2$\\
    $\nu_4\rightarrow\nu_\mu \pi^0$&
    $f_\pi^2/32\pi\left(1-m_\pi^2/m_4^2\right)^2$ & $139.6$\\
    $\nu_4\rightarrow \nu_{\mu}\mu^+\mu^-$ & $\mathrm{Neglected}$ & $211.2$\\
    $\nu_4\rightarrow\mu^{-} \pi^{+}$ & $
    \begin{multlined}\\
    |V_{u\bar{d}}|^2f_\pi^2/32\pi \\
    \left(\left(1-m_\mu^2/m_4^2\right)^2-m_\pi^2/m_4^2\left(1+m_\mu^2/m_4^2\right)\right) \\
    \sqrt{\left(1-(m_\pi^2+m_\mu^2)/m_4^2\right)^2-4m_\pi^2m_\mu^2/m_4^4}
    \end{multlined}
    $ & $245.3$\\
    \hline
    \end{tabular}
\caption{Decay channels up to $m_4\lesssim 250~\rm{MeV}$, for a $\nu_4$ mixed with $\nu_{\mu}$ where $f_\pi =135~\rm{MeV}$, $\tilde{g}_L=-\frac{1}{2}+ \sin^2 \theta_W$, $g_R=\sin^2 \theta_W$, and the electron mass is neglected~\cite{Mastrototaro:2019vug,Gorbunov:2007ak}. The decay mode into two muons is neglected since it is characterized by a small branching ratio for $m_4<500~\mathrm{MeV}$, as shown in Fig.~\ref{fig:BR}. The decay processes for the sterile neutrinos mixed with $\nu_\tau$ via $|U_{\tau4}|^2$ are the ones not involving a single muons in the final states, i.e. all but $\nu_4\to \nu_e\,e^+\,\mu^-$ and $\nu_4\rightarrow \mu^-\,\pi^+$.}
\label{decayt}
\end{table*}

Sterile neutrinos with sufficiently strong interactions with ordinary matter are trapped in the SN via interactions with active neutrinos, electrons, positrons and neutrons. The considered processes are listed in Tab.~\ref{tab:scatteringt}. Moreover, depending on their mass, 
sterile neutrinos may decay in different particles after their production, through the processes shown in Tab.~\ref{decayt}. In Fig.~\ref{fig:BR}, we show the branching ratios for the relevant decay channels. When evaluating the $\nu_{4}$ absorption mfp, we have to consider absorptions and decays separately. In the former case, the mfp is defined as
\begin{equation}
    \lambda_{\rm abs}^{-1} (E_4)=n\,\sigma (E_4)
    \label{eq:lambdas}\,
\end{equation}
where $n$ is the density of targets and $\sigma$ is total absorption cross-section. Since all absorption processes in Tab.~\ref{tab:scatteringt} are $2\to2$ scatterings, it is possible to write the following expression for the cross section
\begin{equation}
\begin{split}
\sigma(E_4)&=\frac{1}{n}\frac{1}{2\,p_4}\int \frac{ {\rm d}^3p_1}{2E_1(2\pi)^3}\frac{{\rm d}^3p_2}{2E_2(2\pi)^3}\frac{{\rm d}^3p_3}{2E_3(2\pi)^3}\\
    &(2\pi)^4 \delta^{4}(p_1+p_2-p_3-p_4) |\mathcal{M}|^2 \\ &f_{3}(1-f_{1})(1-f_{2})\,\, ,
    \end{split}
\label{eq:crossect}
\end{equation}
with a suitable choice of Fermi-Dirac distribution functions $f_{i}$ for $i=1,\,2,\,3$, and of the matrix element $|\mathcal{M}|^{2}$, taken from Tab.~\ref{tab:scatteringt}. In this context, the mfp can be explicitly evaluated by employing the procedure discussed in Ref.~\cite{Hannestad:1995rs}.

Regarding the sterile neutrino decays, the mfp is defined as 
\begin{equation}
\lambda_{\rm dec}=\frac{\gamma v}{\Gamma_{\mathrm{tot}}} \,\ ,
\label{eq:decayels}
\end{equation}
where $\Gamma_{\mathrm{tot}}$ is the sum of the decay widths $\Gamma$ of all the relevant decay processes (see Tab.~\ref{decayt}), $\gamma=(1-\beta^2)^{-1/2}$ is the Lorentz factor and the $\nu_{4}$ velocity is $\beta=p/E_4$. Note that Dirac neutrinos are implicitly assumed throughout our paper; for Majorana states, the rates for exclusive processes are the same, but $L$-conjugated processes, e.g. in decays, are also allowed, thus doubling the inclusive rates. 

By combining Eq.~\eqref{eq:lambdas} and Eq.~\eqref{eq:decayels}, we obtain the total mfp as
\begin{equation}
    \lambda^{-1}=\lambda_{\rm abs}^{-1}+\lambda_{\rm dec}^{-1} \,\ ,
\end{equation}
which is used to evaluate the sterile neutrino luminosity in Eq.~\eqref{eq:la} and impose the constraint in Eq.~\eqref{eq:constraintcooling}. Here it is important to mention that for an unstable particle, the absorption factor in Eq.~\eqref{eq:absfac} would always be zero if the integration limit in the exponential were infinity (since unstable particles can decay in vacuum). However it is possible to fix the upper integration limit in Eq.~\eqref{eq:absfac} to a far radius $R_{\rm far}=100~\km$~\cite{Chang:2016ntp,Chang:2018rso,1993ApJ...412..192B}, much larger than the protoneutron star radius $R_{\rm PNS}\approx 10~\km$ where sterile neutrinos are produced. This choice for $R_{\rm far}$ is an order-of-magnitude estimation for the neutrino gain radius~\cite{1993ApJ...412..192B}. In this way, we assume that only sterile neutrinos reaching the gain radius contribute to the SN cooling. Uncertainties related to the choice of $R_{\rm far}$ will not affect the conclusions of this work, since the cooling bound is always subdominant with respect to other constraints, as discussed in the following.

\begin{figure}[t!]
    \centering
    \includegraphics[width=1.\columnwidth]{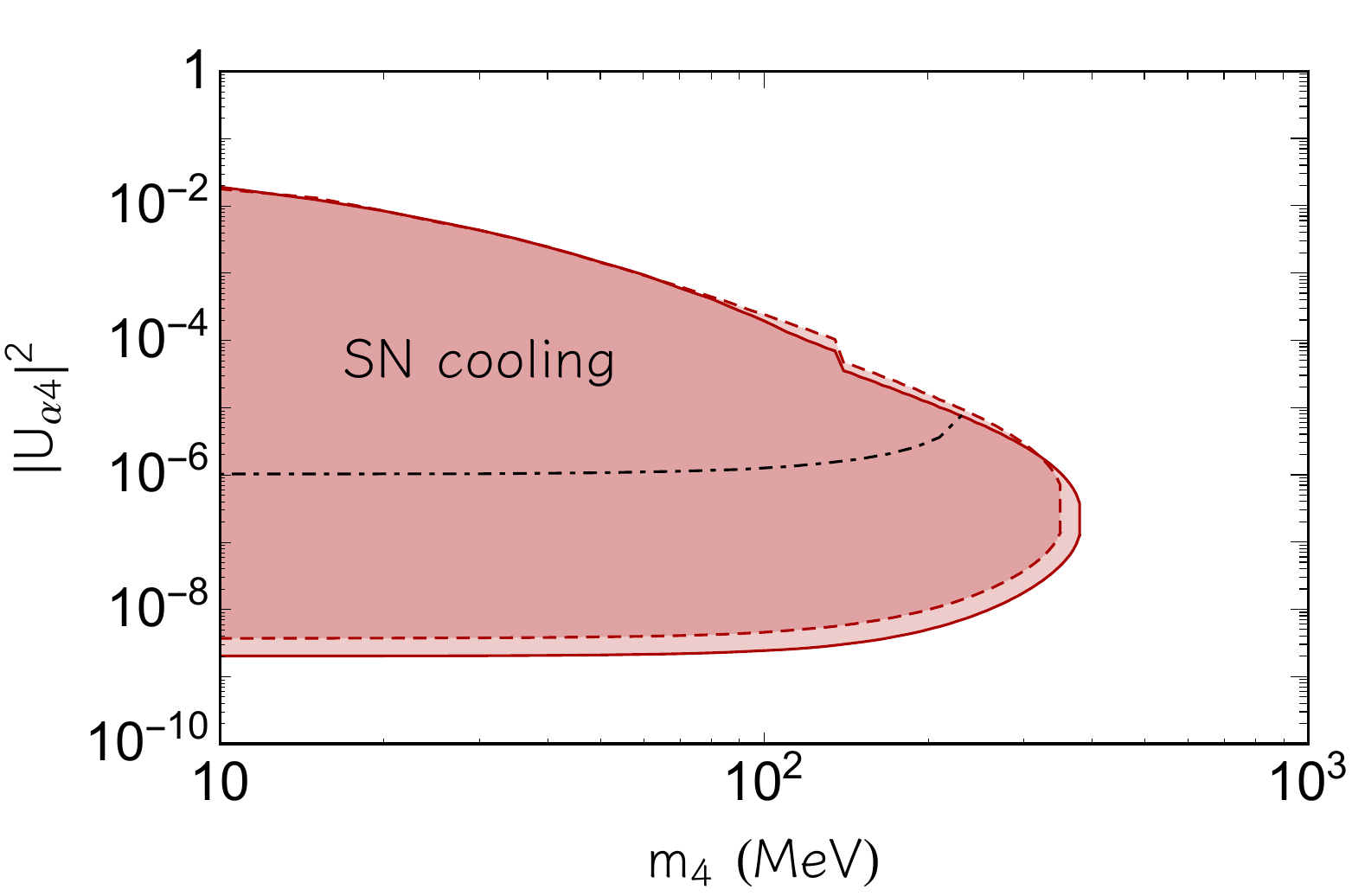}
    \caption{Cooling bound on the sterile neutrino parameter space $(m_4,|U_{\alpha 4}|^2)$ for $\alpha=\mu$ (solid line) and $\alpha=\tau$ (dashed line). The dot-dashed black line is the bound for $|U_{\tau 4}|^2$ which would be obtained neglecting the $\nu N\rightarrow N \nu_4$ process as in Ref.~\cite{Mastrototaro:2019vug}.
    }
    \label{fig:SNCooling}
\end{figure}

The obtained bound is shown in Fig.~\ref{fig:SNCooling}. The contour area delimited by the solid line refers to the mixing with $\nu_\mu$, while the dashed line represents the bound for $\nu_\tau$ mixing. This criterion excludes a region between $\mathcal{O}(10^{-9})<|U_{\alpha 4}|^2<\mathcal{O}(10^{-2})$ for $m_4\sim\mathcal{O}(10)~\mathrm{MeV}$, probing masses up to $\sim 400$~MeV for $|U_{\alpha 4}|^2\sim\mathcal{O}(10^{-7})$. 
Our bound agrees at the order of magnitude level with the bound estimated in the seminal papers~\cite{Dolgov:2000pj,Dolgov:2000jw}.
In the same figure, the dot-dashed black line
shows the lower bound obtained neglecting the $\nu\, N\rightarrow N\,\nu_4 $ interaction. Notice that this constraint is in agreement with the one obtained in Ref.~\cite{Mastrototaro:2019vug} under the same assumption~\footnote{ As a corollary, the magnitude of the signatures associated to benchmarks discussed in~\cite{Mastrototaro:2019vug} remain roughly the same if the benchmark mixings adopted there are scaled-down by the same amount as the tightening of the bounds in presence of neutral current interactions with nuclei.}. Thus, we confirm the crucial importance of the inclusion of neutral interaction processes with nucleons in obtaining the lower exclusion bound on $|U_{\alpha 4}|^2$.
Let us also highlight that the bound for $\alpha=\mu$ is a factor $\sim 2$ stronger than $\alpha=\tau$ because of the larger sterile neutrino luminosity caused by the presence of muons. Henceforth, bounds in the literature ignoring this effect tend to be too conservative. Note that in principle, following a calculation similar to the one in~\cite{Mastrototaro:2019vug}, a related bound might be obtained by the non-observation of a high-energy neutrino flux, from sterile neutrino decay, in the experiments that detected the neutrino signal from the SN 1987A. Our estimates suggest that while  comparable or slightly better than the cooling bound, this would not be competitive with other constraints discussed below, and will not be considered further in this article.

\subsection{SN explosion energy bound}\label{sec:expl_en_bound}

As we can see from Tab.~\ref{decayt}, all the sterile neutrino decay channels except for $\nu_4\rightarrow\nu_\mu\nu_\alpha\bar{\nu}_\alpha$ (with $\alpha=e,\mu,\tau$) produce photons, leptons or pions. If sterile neutrinos have a decay length between the core radius of about 10~km and the progenitor star radius of about $10^{13}~\cm$, they decay inside the SN envelope, depositing at least part of their energy inside the star. This phenomenon allows us to use SNe as efficient calorimeters. As proposed in Refs.~\cite{Sung:2019xie,Caputo:2022mah}, there is an upper limit on the amount of energy that can be deposited inside a SN by FIP decays without producing too energetic explosions that would be incompatible with observations of low-energy SNe.
This constraint requires that
\begin{equation}
E_{\rm FIP}^{\rm e.m.} \lesssim 10^{50}~\mathrm{erg} \,\ , 
\label{eq:lumenX_1}
\end{equation}
where $E_{\rm FIP}^{\rm e.m.}$ is the energy released in the electromagnetic sector by sterile neutrino decays. 

\begin{figure}[t!]
    \centering
    \includegraphics[width=1.\columnwidth]{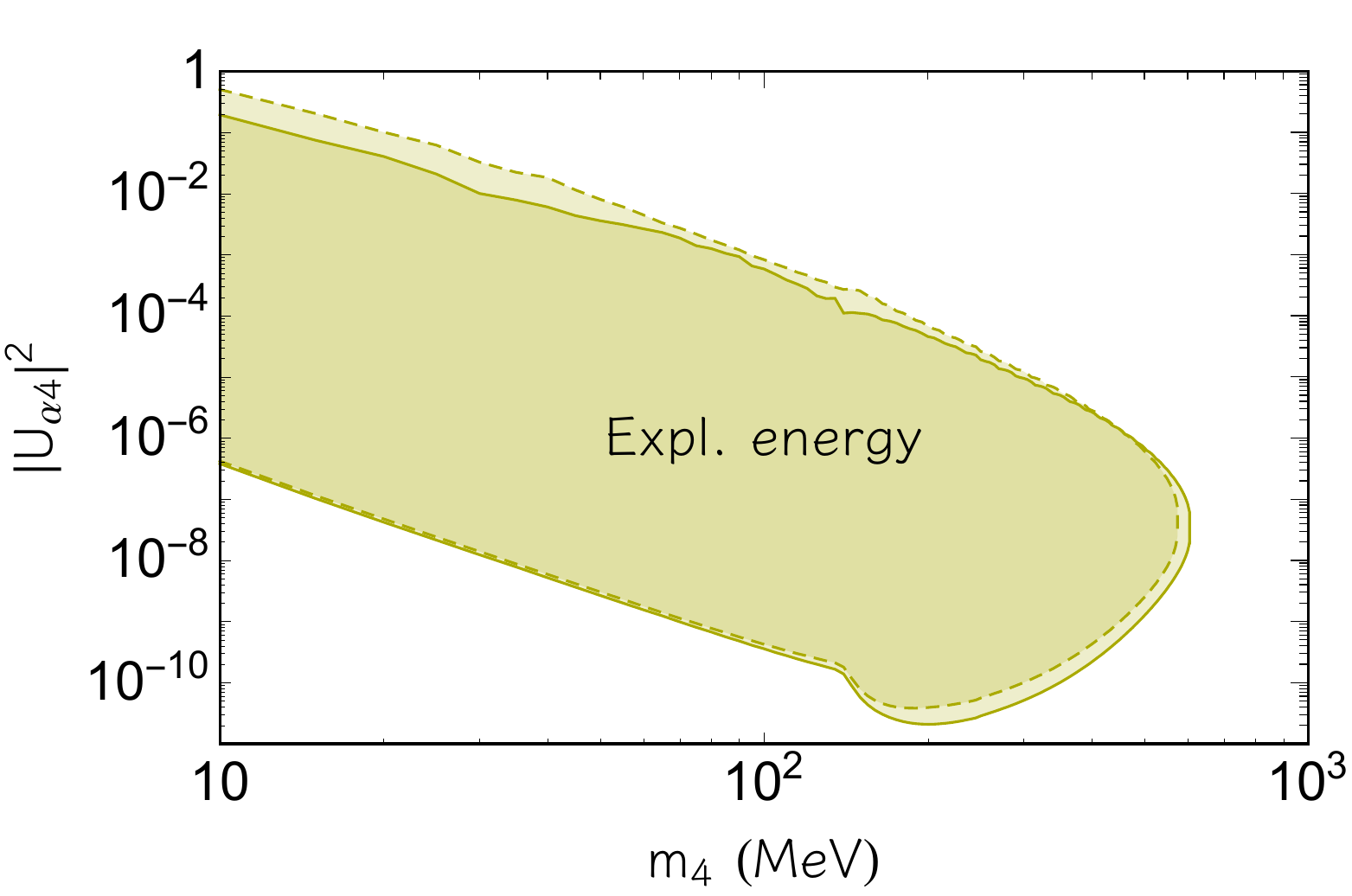}
    \caption{Explosion energy bound on the sterile neutrino parameter space $(m_4,|U_{\alpha 4}|^2)$ for $\alpha=\mu$ (solid line) and $\alpha=\tau$ (dashed line).
    }
    \label{fig:ExpEnergy}
\end{figure}

In the  decays, we assume that the daughter particles are emitted with an appropriate fraction of the energy in the center-of-mass frame, depending on the channel. Following Ref.~\cite{Mastrototaro:2019vug}, it is possible to write the deposited energy as
\begin{equation}
\begin{split}
   E_X^{\rm e.m.}&=\sum_i B_i\frac{m_4}{2\bar{E}}\int{\rm d}t \int {\rm d}E E \int_{E_4^{\mathrm{min}}}^\infty {\rm d}E_4\\
   &\quad\quad\frac{1}{p_4}\frac{{\rm d}^2N_4}{{\rm d}E_4{\rm d}t} \left(1-e^{-(R^*-R_{\rm p})/\lambda_{\rm dec}}\right) \,\ ,
  \end{split}
\end{equation} 
where the index $i$ runs over the decay processes under consideration, $B_i$ is the branching ratio of the $i$-th~process, $\bar{E}$ and $E$ are the daughter particle energies in the center-of-mass and in the laboratory frame, respectively, $E_{4}$ is the sterile neutrino energy, $R^*=2.5\times 10^{13}~\cm$ is the stellar radius and~\cite{Mastrototaro:2019vug,Caputo:2022mah}
\begin{equation}
\begin{split}
E_4^{\mathrm{min}}&=m_4\frac{E^2+\bar{E}^2}{2E\bar{E}} \,\ , \\
\frac{{\rm d}^2N_4}{{\rm d}E_4{\rm d}t}&=4\pi \int_0^{R_{\rm p}} {\rm d}r\, r^2\, \alpha^2(r) \frac{{\rm d}^2{n}_{4}}{{\rm d} E_4{\rm d}t} \langle e^{-\tau(E_4',r)}\rangle\,,
\end{split}
\label{eq:dNdE}
\end{equation}
with ${\rm d}^2N_4/{\rm d}E_4{\rm d}t$ accounting for the fraction of sterile neutrinos escaping from the core, with $R_{\rm p}=40~\km$. In this way, we take into account only the energy carried out from the core and deposited by sterile neutrinos decaying in the SN envelope. We expect the bound in Eq.~\eqref{eq:lumenX_1} to set a constraint on $|U_{\alpha4}|^{2}$ two orders of magnitude more stringent than the SN cooling bound, for sufficiently high $\nu_{4}$ masses. At lower masses,  the longer lifetime and larger boost factors imply that decays are not efficient and this constraint is relaxed.
Indeed, in Fig.~\ref{fig:ExpEnergy}
we see that for ${m_4>\mathcal{O}(100)~\mathrm{MeV}}$, the bound excludes values of the mixing down to $|U_{\alpha 4}|^2\sim \mathcal{O}(10^{-10})$. The bump in the constraint around $135$~MeV reflects the opening of an extra decay channel for sterile neutrinos, $\nu_{4}\to\nu_{\alpha}\pi^{0}$.
Recently, the authors of Ref.~\cite{Chauhan:2023sci} have obtained a SN explosion energy bound without considering the $\nu N\rightarrow\nu_4 N$ interaction. Similarly to the SN cooling case, neglecting the neutral current interactions with nucleons leads to a constraint two orders of magnitude weaker than the one obtained in Fig.~\ref{fig:ExpEnergy}.

\subsection{511 keV bound}
\label{sec:511 keV bound}

Sterile neutrinos escaping the SN envelope and decaying in the interstellar medium give rise to a diverse phenomenology, depending on the considered decay products. Here, we focus on the positrons produced by a portion of the $\nu_{4}$-decay channels.

As extensively discussed in Refs.~\cite{Calore:2021klc,Calore:2021lih,DelaTorreLuque:2023nhh,DelaTorreLuque:2023huu}, this exotic injection of positrons in the Galaxy would originate a distinctive soft gamma-ray signal. Precisely, positrons emitted by sterile neutrino decays are trapped in the Galaxy by its magnetic field. While traveling on scales smaller than $\mathcal{O}(1)~\kpc$ from the decay point, positrons lose energy by Bhabha scattering on the galactic electron population. This thermalization process lasts between $10^{3}$ and $10^{6}$~yrs, depending on the electron density. This long time-scale explains why the positron injection, caused by SNe during the history of the Galaxy, can be assumed continuous.
Once positrons are almost at rest, $\sim 25\%$ of them form a parapositronium bound state with an electron, before decaying in two back-to-back photons, each one with an energy of $511$~keV, determined by the electron rest mass~\cite{Karshenboim:2003vs}. 

\begin{figure}[t!]
    \centering
    \includegraphics[width=1.\columnwidth]{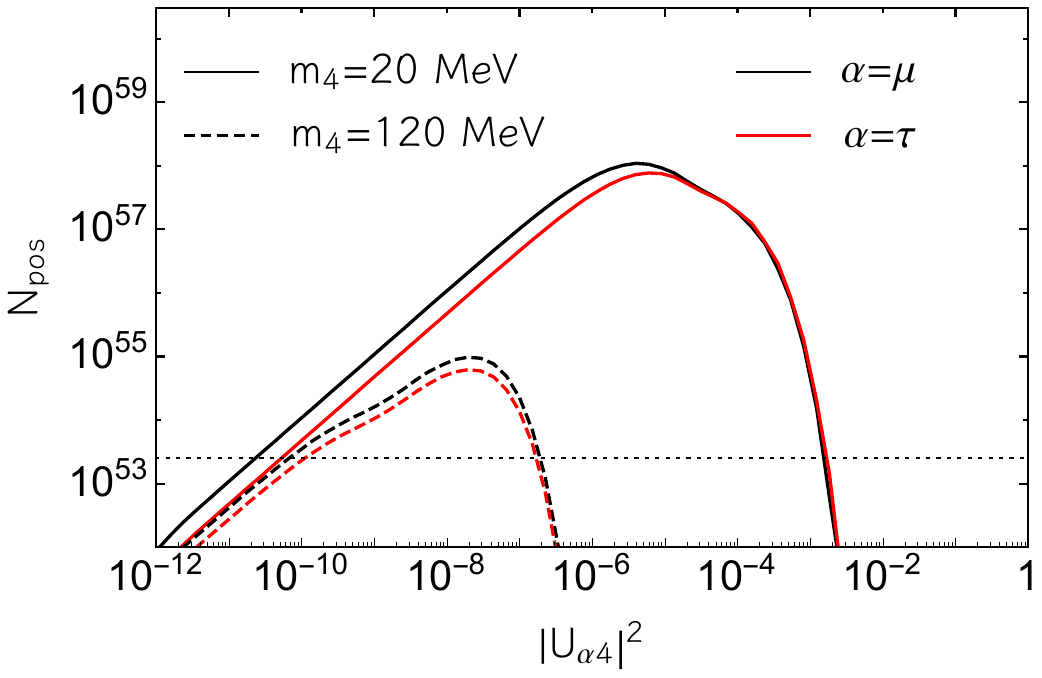}
    \caption{Number of produced positrons per SN for $\alpha=\mu,\tau$ (black and red lines, respectively) and $m_4=20~\mathrm{MeV}$ (solid lines) and $m_4=120~\mathrm{MeV}$ (dashed lines). The dotted line corresponds to the limit value of $N_{\rm pos}=2.5\times10^{53}$. 
    }
    \label{fig:N}
\end{figure}

A Galactic $511$~keV line, at least partially explained by standard positron emission mechanisms,  is prominently observed from the direction of the Galactic bulge~\cite{Prantzos:2010wi}.  The contribution to this signal induced by sterile neutrinos can be calculated as
\begin{equation}
\frac{{\rm d} \phi_\gamma^{ 511}}{{\rm d}\Omega}= 2k_{ps}N_{\rm pos} \Gamma_{cc}\int {\rm d}s \, \frac{n_{cc}[r(s,b,l),z(s,b)]}{4 \pi } \;,
\label{eq:phot}
\end{equation}
where ${\rm d}\Omega = {\rm d}l\, {\rm d}b \cos b$, with $- \pi \leq l \leq \pi$ being the longitude and $-\pi/2 \leq b \leq \pi/2$ being the latitude in the Galactic coordinate system $(s,b,l)$, with $s$ distance from the SN to the Sun. Moreover, ${k_{ps}=1/4}$ accounts for the fraction of positrons annihilating through parapositronium. According to Ref.~\cite{Li:2010kd}, we fix ${\Gamma_{cc}=2.30}$~SNe/century as the Galactic SN rate. Finally, $n_{cc}$ is the SN volume distribution~\cite{Mirizzi:2006xx} in the Galactocentric coordinate system $(r,z,l)$, with $r$ the galactocentric radius and $z$ the height above the Galactic plane, connected with the Galactic coordinate system through the relations
\begin{equation}
\begin{split}
r& = \sqrt{s^2 \cos^2 b + d_{\odot}^2 -2 d_\odot s\cos l \cos b}\,\ , \\ 
z&=s \sin b \,.
\end{split}
\end{equation}
Here, we set  the solar distance from the Galactic center to $d_{\odot}=8.5$ kpc. 
Requiring the photon flux in Eq.~\eqref{eq:phot} to be smaller than the observed signal in the range ${l\in [28.25^{\circ};31.25^{\circ}]}$ and ${b\in [-10.75^{\circ};10.25^{\circ}]}$,
we obtain a constraint on the number of injected positrons~\cite{DelaTorreLuque:2023huu}
\begin{equation}
N_{\rm{pos}}\lesssim 2.5\times 10^{53} \,\ .
\label{eq:nposbound}
\end{equation}
This is the most conservative limit obtained by the comprehensive analyses of Refs.~\cite{DelaTorreLuque:2023huu,DelaTorreLuque:2023nhh}, taking into account different SN distribution models and diffusive smearing effects. This upper bound on $N_{\rm{pos}}$ corresponds also to the constraint placed by XMM-Newton observations of the Galactic X-ray background~\cite{Foster:2021ngm}. Indeed, an excess of electron/positron injection in the Galaxy would source a diffuse X-ray signal via inverse Compton scattering on the stellar background light.

\begin{figure}[t!]
    \centering
    \includegraphics[width=1.\columnwidth]{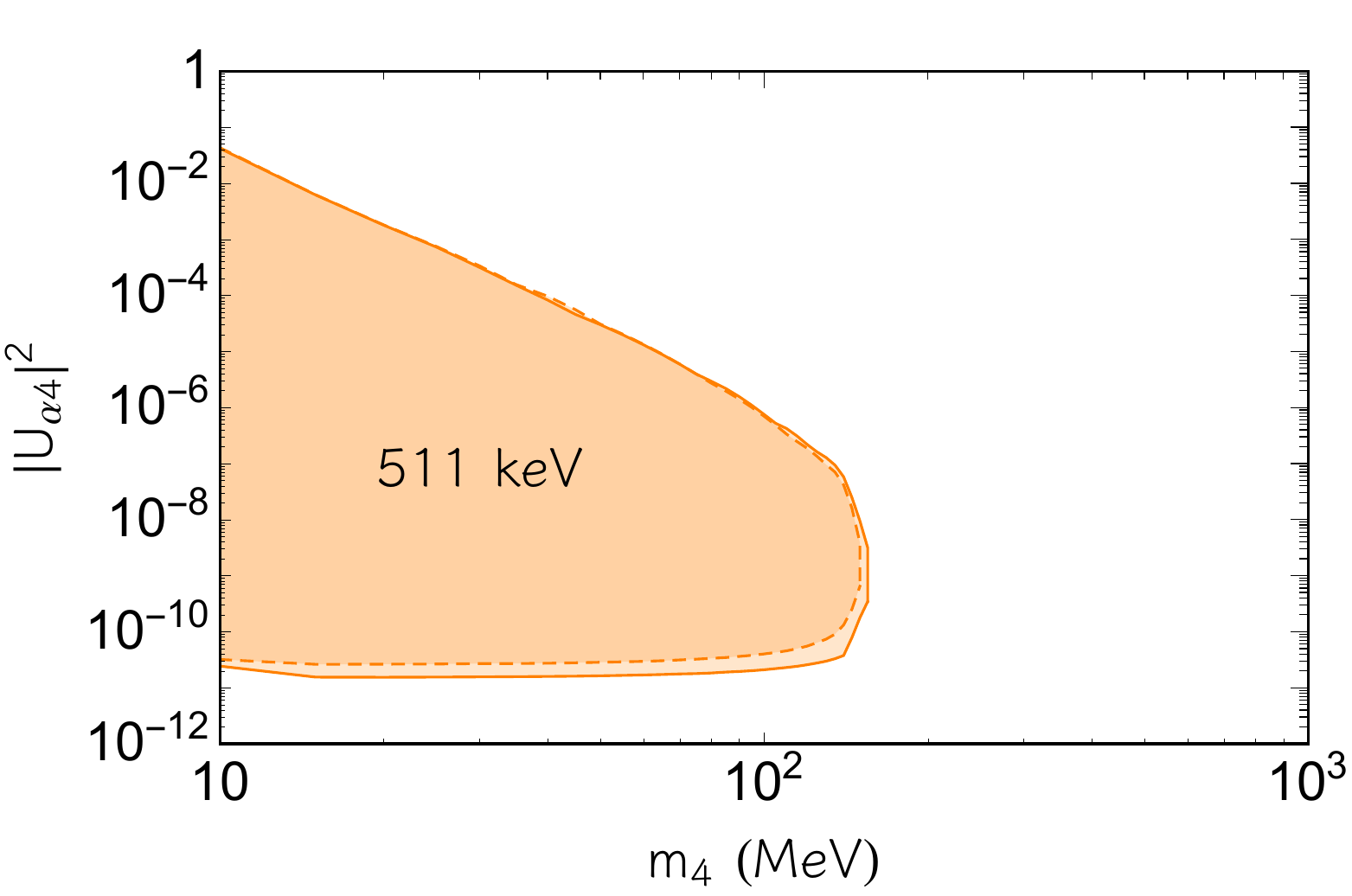}
    \caption{$511~\mathrm{keV}$ line bound on the sterile neutrino parameter space $(m_4,|U_{\alpha 4}|^2)$ for $\alpha=\mu$ (solid line) and $\alpha=\tau$ (dashed line).}
    \label{fig:511}
\end{figure}

In order to apply the constraint in Eq.~\eqref{eq:nposbound}, we calculate the number of injected positrons as 
\begin{equation}
\begin{split}
    N_{\rm pos}& = n_{\rm pos}\int {\rm d}E_4\frac{{\rm d}N_4}{{\rm d}E_4}\left(\epsilon_{II}e^{-{r_{II}}/\lambda_{\rm dec}}+\epsilon_{I}\,e^{-{r_{\rm I}}/\lambda_{\rm dec}}\right)\,,
    \label{eq:npos}
\end{split}
\end{equation}
with
\begin{equation}
\label{eq:averagepositrons}
    n_{\rm pos}=\sum_{i}n_i B_i \,\ ,
\end{equation}
the average number of positrons produced in a sterile neutrino decay. Moreover, following Ref.~\cite{DeRocco:2019njg} we fix
\begin{equation}
    r_{II}=10^{14}~{\rm cm}, \qquad  r_{I}=2\times 10^{12}~{\rm cm}\,,
\end{equation}
for the envelope radii of Type II and Ib/c SNe, while according to Ref.~\cite{Li:2010kd}, we take as average fractions of SNe of Type II and Ib/c 
\begin{equation}
    \epsilon_{II}=1-\epsilon_{I},\qquad \epsilon_{I}=0.33\,\ .
\end{equation}
In Fig.~\ref{fig:N} we show the calculated $N_{\rm pos}$ as a function of the mixing angle. At low mixing, sterile neutrinos are not efficiently produced and, therefore, the number of positrons produced in the decay is smaller than the limiting value, represented by the dotted line. Then, as the sterile neutrino production increases, given that almost the totality of neutrinos decay inside the Galaxy, the injected positrons can be a sizable number. As it can be seen in Fig.~\ref{fig:N}, light sterile neutrinos with $m_{4}=20$~MeV (solid lines) can produce up to $\sim10^{58}$ positrons per SN. A smaller number is obtained by more massive neutrinos (dashed lines), since their production is Boltzmann suppressed. For small values of the mixing (e.g., $|U_{\alpha\,4}|^2\lesssim 10^{-6}$ for $m_4=20~\MeV$), we notice a relatively small difference between sterile neutrinos mixed with muon neutrinos (black lines) or tau neutrinos (red lines), due to the larger production of the former ones induced by charged current interactions of muons with nucleons. For larger values of the mixing (e.g., $|U_{\alpha\,4}|^2\gtrsim 10^{-6}$ for $m_4=20~\MeV$), the number of positrons is exponentially suppressed and the different production and absorption processes lead to an even smaller difference in the positron production. The bound obtained with this approach is expected to exclude relatively light $\nu_{4}$, with masses above a few tens of MeV, and it can be extended to small couplings because it is a cumulative diffuse flux. Indeed, we can see from Fig.~\ref{fig:511} that the obtained lower bound is $|U_{\alpha 4}|^2\sim\mathcal{O}(10^{-11})$ for $m_4<\mathcal{O}(100)~\mathrm{MeV}$. 

\subsection{SN 1987A gamma-ray bound}
\label{sec:SN 1987A gamma-ray bound}

As discussed in the previous subsection, sterile neutrinos decaying after escaping the SN envelope lead to peculiar signatures. One of the most powerful constraints is given by the non-detection of a gamma-ray signal in coincidence with the neutrino burst of SN 1987A, as studied in the seminal work of Ref.~\cite{Oberauer:1993yr}. The Gamma-Ray Spectrometer of the Solar Maximum Mission places an upper limit of~\cite{OBERAUER1993377} 
\begin{equation}
    \phi_{\gamma}\lesssim1.38 \cm^{-2}\,,
    \label{eq:SMM}
\end{equation}
on the photon flux at energies between 25~MeV and 100~MeV for 232.2~s after the first neutrino arrival. This upper limit translates into a constraint on $|U_{\alpha 4}|^{2}$, since the radiative decay of massive $\nu_{4}$ would give rise to a gamma-ray signal in coincidence with a SN explosion.
From Tab.~\ref{decayt}, we notice that the only decays contributing to this signal are $\nu_{4}\to \nu_{\alpha} \gamma$ and $\nu_{4}\to \nu_{\alpha}\pi^{0}$, because of the successive decay $\pi^{0}\to \gamma\gamma$. The spectrum of photons originated by $\nu_{4}$ decay directly into photons is~\cite{Oberauer:1993yr} 
\begin{equation}
     \left(\frac{{\rm d} N_{\gamma}}{{\rm d}E_{\gamma}}\right)_{\rm dir}=\frac{m_4}{2\bar{E}_{\nu_4\rightarrow\nu_\alpha\gamma}}B_{\nu_4\rightarrow\nu_\alpha\gamma}\int_{E^{\rm{min}}_{\nu_4\rightarrow\nu_\alpha\gamma}}^{\infty}{\rm d}E_4\frac{1}{p_4}\frac{{\rm d}N_4^{\rm{esc}}}{{\rm d}E_4}\,,
     \label{eq:gammadecays}
\end{equation}
where the average energy, in the center-of-mass frame, of the daughter particle $j$, from the decay of the parent particle $i$ is
\begin{equation}
    \begin{split}
    \bar{E}_{i\rightarrow j}&=\frac{m^2_i-m^2_j}{2m_i} \,\ ,
    \end{split}
\end{equation}
which is larger or equal to
\begin{equation}
    \begin{split}
         E^{\rm{min}}_{i\rightarrow j}&=m_i\frac{E_i^2+\bar{E}_{i\rightarrow j}^2}{2E_i\bar{E}_{i\rightarrow j}} \,\ ,
    \end{split}
\end{equation}
when expressed in the laboratory frame.
In addition, the fraction of sterile neutrinos decaying outside the SN envelope is
\begin{equation}
     \frac{{\rm d}N_4^{\rm{esc}}}{{\rm d}E_4}=\frac{{\rm d}N_4}{{\rm d}E_4}e^{-R^*/\lambda_{\rm dec}} \,.
\end{equation}

Similarly to Eq.~\eqref{eq:gammadecays}, we can evaluate the energy spectrum of neutral pions produced by sterile neutrinos as
\begin{equation}
     \frac{{\rm d}N_{\pi^{0}}}{{\rm d}E_{\pi^{0}}}=\frac{m_4B_{\nu_4\rightarrow\nu_\alpha\pi^{0}}}{2\bar{E}_{\nu_4\rightarrow\nu_\alpha\pi^{0}}}\int_{E^{\rm{min}}_{\nu_4\rightarrow\nu_\alpha\pi^{0}}}^{\infty}\frac{{\rm d}E_4}{p_4}\frac{{\rm d}N_4^{\rm{esc}}}{{\rm d}E_4}\,.
     \label{eq:piondecays}
\end{equation}
In a second step, the gamma-ray spectrum from the almost immediate pion decay is obtained as~\cite{Oberauer:1993yr}
\begin{equation}
   \left(\frac{{\rm d}N_{\gamma}}{{\rm d}E_{\gamma}}\right)_{\rm \pi^{0}}= \frac{m_{\pi^0}}{2\bar{E}_{\pi^0\rightarrow\gamma\gamma}}\int_{E^{\rm{min}}_{\pi^0\rightarrow\gamma\gamma}}^\infty \frac{{\rm d}E_{\pi^0}}{p_{\pi^0}}\frac{{\rm d}N_{\pi^0}}{{\rm d}E_{\pi^0}} \,\ .
\end{equation}
In conclusion, the expected gamma-ray flux can be written as
\begin{equation}
    \frac{{\rm d}\phi_\gamma}{{\rm d}E_\gamma {\rm d}t}=\frac{1}{4\pi d^2_{\rm SN}}\frac{\beta e^{-t\beta/\lambda_{\rm dec}}}{\lambda_{\rm dec}}\frac{{\rm d}N_{\gamma}}{{\rm d}E_{\gamma}}\,,
    \label{eq:grbflux}
\end{equation}
where
\begin{equation}
    \frac{{\rm d}N_{\gamma}}{{\rm d}E_{\gamma}}=\left(\frac{{\rm d}N_{\gamma}}{{\rm d}E_{\gamma}}\right)_{\rm dir}+\left(\frac{{\rm d}N_{\gamma}}{{\rm d}E_{\gamma}}\right)_{\rm \pi^{0}}\, .
\end{equation}

\begin{figure}[t!]
    \centering
    \includegraphics[width=1.\columnwidth]{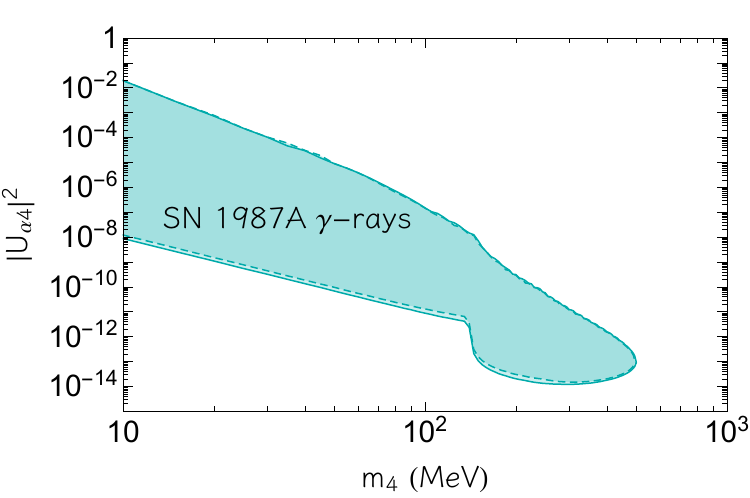}
    \caption{Bound on the sterile neutrino parameter space $(m_4,|U_{\alpha 4}|^2)$ from the non-detection of  gamma-rays from SN 1987A for $\alpha=\mu$ (solid line) and $\alpha=\tau$ (dashed line).}
    \label{fig:SNray}
\end{figure}

We set stringent constraints on the sterile neutrino properties by integrating Eq.~\eqref{eq:grbflux} over the observation time of 232.2~s and comparing the result with the limit in Eq.~\eqref{eq:SMM}. The constraint obtained in this way becomes particularly relevant as soon as $\nu_{4}$ is heavier than the pion, opening up the pion decay channel. Our results are reported in Fig.~\ref{fig:SNray}, showing that the lower bound strengthens from $|U_{\alpha 4}|^2\gtrsim \mathcal{O}(10^{-10})$ for $m_4<m_\pi$, $m_\pi$ being the pion mass, to $|U_{\alpha 4}|^2\gtrsim \mathcal{O}(10^{-12})$ for $m_4>m_\pi$ due to the decay channel $\nu_4\rightarrow\nu_\alpha \pi$. As argued in~\cite{Diamond:2023scc} for axionlike particles, we mention that for sterile neutrinos with masses of a few 10 MeV the decay-product photons may create a fireball, making part of the ``SN 1987A gamma-rays'' bound not valid. However, the fireball would produce a gamma-ray flux with energy of a few MeV and the non-detection of such a signal in coincidence with the SN 1987A burst by Pioneer Venus Orbiter (PVO), constrains again this region, which in turn is already excluded by other bounds such as the 511 keV one.

\subsection{Diffuse gamma-ray bound}\label{sec:diffuse_gamma_bound}
The same phenomenology discussed above can be applied to evaluate the cumulative gamma-ray flux induced by SN $\nu_4$ during the history of the Universe. This would constitute a diffuse, isotropic and constant gamma-ray flux at a few tens of MeV.

\begin{figure}[t!]
    \centering
    \includegraphics[width=1\columnwidth]{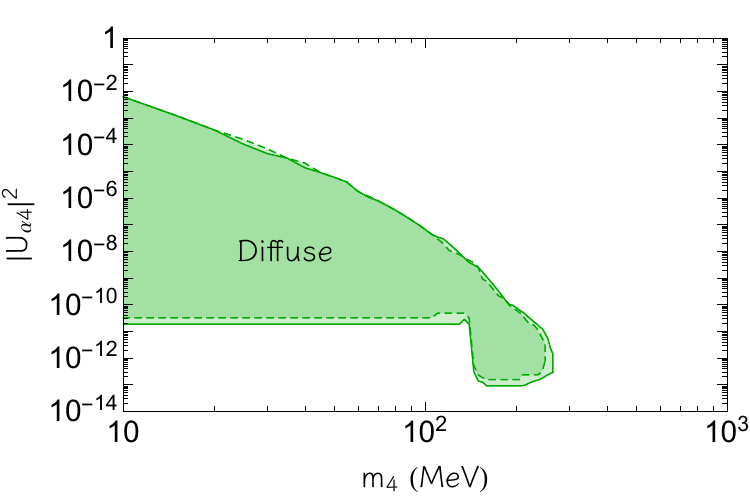}
    \caption{Bound on the sterile neutrino parameter space $(m_4,|U_{\alpha 4}|^2)$ from diffuse gamma-ray emission for $\alpha=\mu$ (solid line) and $\alpha=\tau$ (dashed line).}
    \label{fig:Diffuse}
\end{figure}

The gamma-ray spectrum for a single SN is calculated as in Eq.~\eqref{eq:grbflux}, redshifted in energy and integrated over the SN explosion rate as (see Ref.~\cite{Beacom:2010kk} for calculation details in the case of the SN diffuse neutrino spectrum and Ref.~\cite{Calore:2021hhn} for the axion case)
\begin{equation}
\begin{split}
    \frac{{\rm d}\Phi_{\gamma}}{{\rm d}E_\gamma}=\int_0^\infty &(1+z)\frac{{\rm d}N_\gamma(E_\gamma(1+z))}{{\rm d}E_\gamma} R_{\rm{SN}}(z)\left|\frac{{\rm d}t}{{\rm d}z}\right|{\rm d}z \,\ ,
\end{split}
\label{eq:fluxz}
\end{equation}
where $z$ is the redshift, $R_{\rm{SN}} (z)$ is the SN explosion rate taken from~\cite{Priya:2017bmm}, with a total normalization for the core-collapse rate $R_{\rm{cc}} = 1.25\times 10^{-4} \rm{yr^{-1} Mpc^{-3}}$. Furthermore, $|{\rm d}t/{\rm d}z|^{-1}= H_0(1+z)[\Omega_{\Lambda}+\Omega_M(1+z)^3]^{1/2}$ with
the cosmological parameters fixed at $H_0=67.4 \rm{km~s^{-1}~Mpc^{-1}}$, $\Omega_M = 0.315$, $\Omega_{\Lambda}=0.685$~\cite{Planck:2018vyg}.
The flux in Eq.~\eqref{eq:fluxz} is imposed to be smaller than
\begin{equation}
\frac{{\rm d}\Phi_{\gamma}^{\rm{obs}}}{{\rm d}E_\gamma}=2.2\times 10^{-3}\left(\frac{E}{\mathrm{MeV}}\right)^{-2.2}~\mathrm{MeV^{-1}~cm^{-2}~s^{-1}~sr^{-1}} \,,
\end{equation}
extracted from measurements of {\it Fermi}-LAT of the diffuse gamma-ray background~\cite{Calore:2020tjw}.
The advantage of this constraint, reported in Fig.~\ref{fig:Diffuse}, is that it extends to smaller masses, where the decay rate is less efficient, excluding $|U_{\mu 4}|^2\gtrsim 2\times 10^{-11}$ and $|U_{\tau 4}|^2\gtrsim 3\times 10^{-11}$ for $m_4\lesssim 100~\MeV$.

\section{Combination of different bounds}
\label{sec:comparison}

\begin{figure*}[t!]
    \centering
    \includegraphics[width=2.\columnwidth]{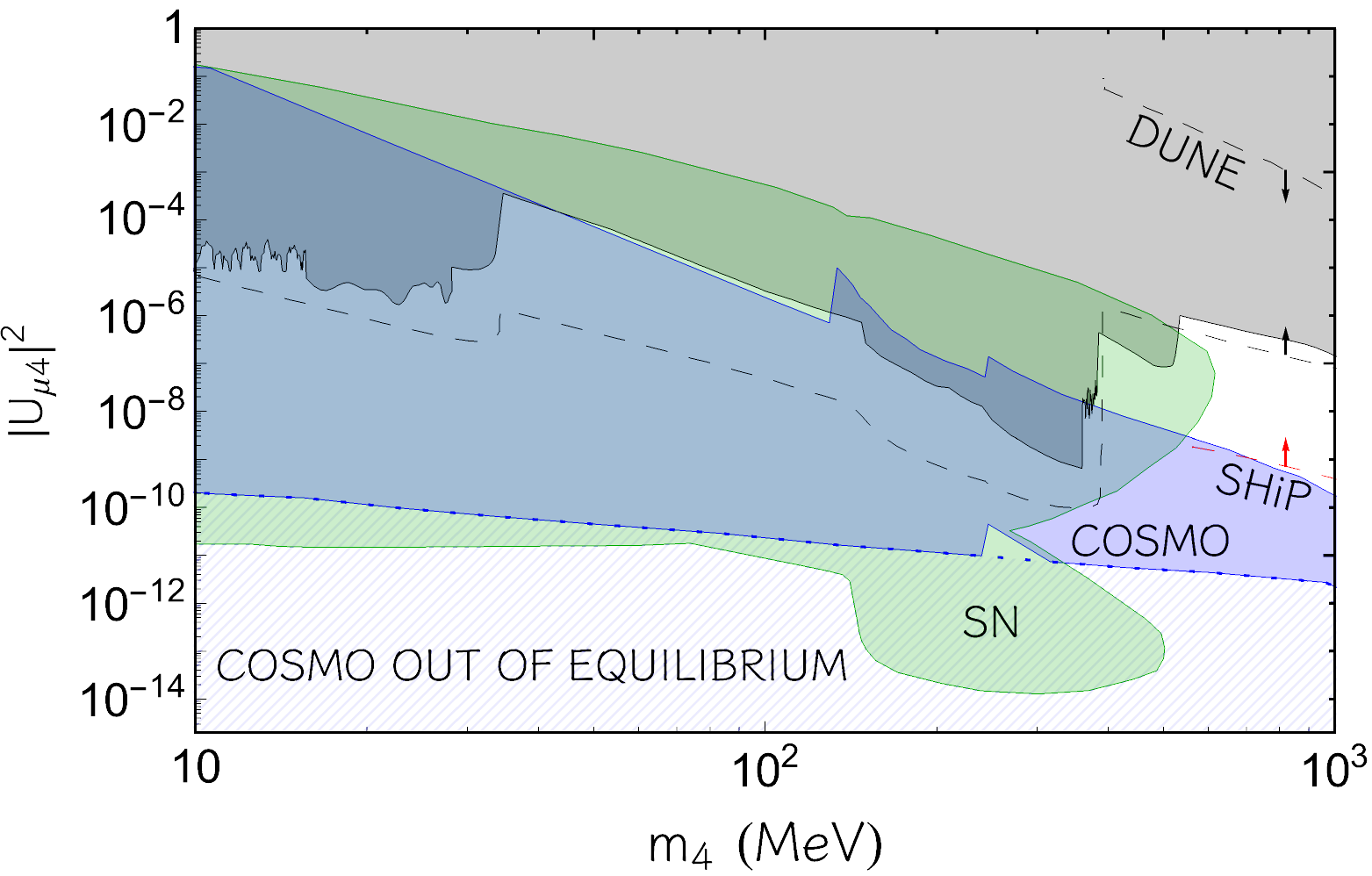}
    \includegraphics[width=2.\columnwidth]{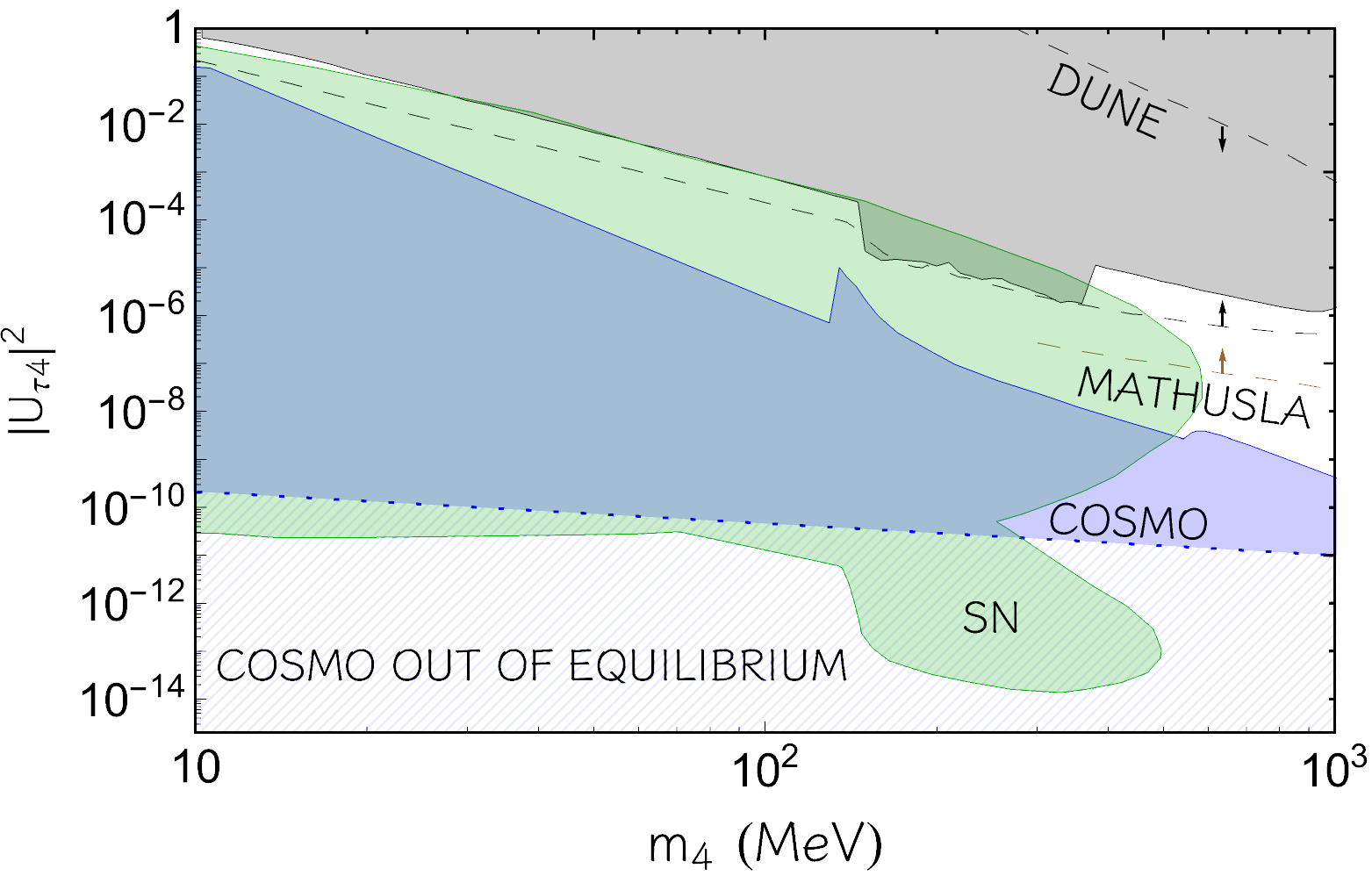}
    \caption{Overview of the bounds from SNe (green region), cosmology~\cite{Sabti:2020yrt,Boyarsky:2020dzc,Mastrototaro:2021wzl} (blue region) and experiments~\cite{Bolton:2019pcu} (gray region) for sterile neutrinos mixed with muon neutrinos (upper panel) and tau neutrinos (lower panel). The dashed lines represent the sensitivities of the future experiments DUNE~\cite{Ballett:2019bgd,Krasnov:2019kdc} (black), SHiP~\cite{SHiP:2018xqw} (red) and MATHUSLA~\cite{Chou:2016lxi} (brown). The hatched area below the dotted blue line represents the region of the parameter space in which sterile neutrinos are produced in the early universe out of equilibrium, whose viability is more model-dependent.}
    \label{fig:compare_bound}
\end{figure*}

In Fig.~\ref{fig:compare_bound} we combine all the bounds obtained in the previous Sections
for $\nu_4$ mixed with $\nu_\mu$ (upper panel)
and $\nu_\tau$ (lower panel). 

\subsection{Laboratory bounds}

Constraints on heavy sterile neutrinos decaying into leptons and pions are set by the long-baseline neutrino oscillation experiment T2K~\cite{T2K:2019jwa}. A beam of 30~GeV protons produced a large amount of kaons in their scattering on a graphite target at J-PARC. Then, kaons might produce sterile neutrinos in their decay. A detector placed at a baseline of 280 m was used to reveal the decay of sterile neutrinos. The constraints obtained by T2K complement and improve the results of CHARM~\cite{CHARMII:1994jjr} and PS191~\cite{Bernardi:1987ek}. Current experimental bounds are shown as the gray shaded area in Fig.~\ref{fig:compare_bound}. 

The aforementioned searches will be improved by future experiments, whose projected sensitivities are shown as dashed lines in Fig.~\ref{fig:compare_bound}. In particular, current experimental constraints on sub-GeV sterile neutrinos considered in this work will be strengthened by DUNE~\cite{Ballett:2019bgd,Krasnov:2019kdc}, probing however regions not excluded by SN arguments only for masses $m_4\gtrsim 400~\MeV$, as shown by the dashed-black line in Fig.~\ref{fig:compare_bound}. Moreover, the future beam-dump experiment SHiP~\cite{Alekhin:2015byh} is designed to probe exotic long-lived particles produced by a 400 GeV proton beam from the Super Proton Synchrotron at CERN, allowing the exploration of a much larger region of the parameter space for sterile neutrinos mixed with muon neutrinos~\cite{SHiP:2018xqw}, as represented by its sensitivity (dashed-red line) in the upper panel of Fig.~\ref{fig:compare_bound}. On the other hand, as shown by the dashed-brown line in the lower panel of Fig.~\ref{fig:compare_bound}, a currently unexplored region of the parameter space of sterile neutrinos mixed with tau neutrinos will be probed by MATHUSLA~\cite{Chou:2016lxi}, another CERN experiment planned to study sterile neutrinos by searching for displaced vertex signatures near the LHC interactions points.

\subsection{Cosmological bound}
Constraints on heavy sterile neutrinos from cosmological observations emerge considering that their decay, after the active neutrino decoupling, generates extra neutrino radiation and entropy production in the Early Universe. Therefore, they alter the value of the effective number of neutrino species $N_{\rm eff}$, measured by the cosmic microwave background (CMB), and affect primordial nucleosynthesis (BBN), notably ${}^4$He production, which is reflected in the $Y_p$ value. Using the latest measurements of the Planck collaboration~\cite{ParticleDataGroup:2018ovx, Planck:2018vyg}, it is possible to obtain cosmological constraints, see~\cite{Sabti:2020yrt,Boyarsky:2020dzc,Mastrototaro:2021wzl}. These arguments exclude up to $|U_{\alpha 4}|^2\sim\mathcal{O}(10^{-1})$ for $m_4\approx 10~\MeV$, as represented by the blue region in Fig.~\ref{fig:compare_bound} labelled as ``COSMO''. For heavier sterile neutrinos, with mass $m_4>m_\pi$, the strongest impact on BBN is induced by the meson-driven $p \leftrightarrow n$ conversion, which significantly increases the helium abundance and constrains sterile neutrinos with lifetimes larger than $0.02~\s$~\cite{Boyarsky:2020dzc}. The dotted blue line delimiting from below the blue region in Fig.~\ref{fig:compare_bound} corresponds to the limit of validity of the assumptions used to obtain cosmological bounds. Indeed, these constraints are derived by considering only sterile neutrinos thermally produced and sufficiently short-lived so that they do not change the nuclear reaction framework by their meson decay products~\cite{Boyarsky:2020dzc}. In this context, we mention that cosmological constraints may be extended also to the region of the parameter space in which sterile neutrinos are produced non-thermally (the hatched region below the dotted blue line, labelled as ``COSMO OUT OF EQUILIBRIUM''), as discussed for instance in Ref.~\cite{Ovchynnikov:2021zyo}. In this region, however, the bounds are more strongly dependent on the assumptions on the early universe and the extra couplings of these states.
$ $

$ $ 

For $\nu_4$ mixed with $\nu_\mu$, SN arguments lead to the lower limit ${|U_{\mu 4}|^2 \gtrsim \mathcal{O}(10^{-11})}$ for $m_a \lesssim 100$ MeV, notably due to the 511 keV line argument (see Sec.~\ref{sec:511 keV bound}) and the diffuse gamma-ray flux (see Sec.~\ref{sec:diffuse_gamma_bound}). At larger masses, the SN bound tightens to $|U_{\mu 4}|^2 \lesssim 10^{-14}$ in the range $200~\MeV \lesssim m_4 \lesssim 500$ MeV, due to the absence of gamma-rays in coincidence with SN 1987A (see Sec.~\ref{sec:SN 1987A gamma-ray bound}). The existing laboratory bounds nicely complement the SN ones, excluding the parameter space all the way to large mixing angle, and overlapping with SN bounds here dominated by the explosion energy argument (see Sec.~\ref{sec:expl_en_bound}). Future laboratory experiments are expected to charter new parameter space only for $m_4 \gtrsim 500$ MeV, probing mixing angles
 $10^{-6}\lesssim |U_{\mu 4}|^2 \lesssim 10^{-9}$.
In the case of $\nu_4$ mixed with $\nu_\tau$, the situation of the bounds is qualitatively similar. Factor $\sim 2$ differences are due either to the extra production processes for sterile neutrinos associated with charged current interactions with muons, or to the extra decay channels, present only for sterile neutrinos mixed with $\nu_\mu$.

\section{Conclusions}
\label{sec:conclusions}

In this work we revised and improved current bounds on heavy sterile neutrinos mixed with the active ones.  In particular, we considered the cooling bound derived from  neutrino observations from SN 1987A. 
We also studied the decays of heavy sterile neutrinos, affecting the  SN explosion energy and possibly producing a gamma-ray signal.
We improved the characterization of sterile neutrino  neutral current interactions of $\nu_4$  with nucleons. We also include charged current interactions of $\nu_4$  with muons, which is relevant for sterile neutrino production mixing with $\nu_\mu$.
Contrary to consolidate belief, it results that the dominant channel for sterile neutrino production is associated with neutral current interactions. 
Furthermore, we extended the bounds to the 
trapping regime of $\nu_4$ verified at large mixing angles, adopting the the so-called ``modified luminosity criterion''. 
We also strengthened the SN cooling bounds considering (non)radiative decays of heavy  neutrinos, and characterizing their effect 
on excessive energy deposition in the SN envelope,  and the observable gamma-ray signal when decays occur outside the SN.
The combination of all the SN bounds (together with laboratory ones) allows one to exclude values $|U_{\alpha 4}|^2 \gtrsim 2-3\times 10^{-11}$ for $m_a \lesssim 100$ MeV. At larger masses the bound tightens to $|U_{\alpha 4}|^2 \lesssim 10^{-13}-10^{-14}$ in the range $200 \lesssim m_4 \lesssim 500$ MeV. It is worthwhile to mention that another possible astrophysical bound on sterile neutrinos comes from the observation of binary neutron star merger events (see, e.g., Ref.~\cite{Diamond:2023cto} for related constraints on axionlike particles). We reserve this analysis for future work.
The most interesting region that SN bounds leave open for future laboratory searches (such as DUNE, SHiP and MATHUSLA) is the range $10^{-6}\lesssim |U_{\mu 4}|^2 \lesssim 10^{-9}$ for 
$m_4 \gtrsim 500$ MeV.  In the case of mixing with $\nu_\tau$, DUNE would also have the potential to robustly probe the range of masses down to $\sim 10$ MeV at large mixings, where the overlap between SN and laboratory experiments is minimal or absent.

We conclude with two remarks. Having a synoptic view of the bounds following from SN arguments reveals that in most of the parameter space, at least a couple of arguments lead to constraints of similar strength. Since they suffer from different systematics, this is reassuring in supporting the overall reliability of such indirect limits. For instance, diffuse gamma-ray and 511 keV bounds rely on average properties of SN, such as their rate, rather than the single SN 1987A event. Also note that one does not have to rely on the cooling argument, which has been repeatedly criticized in recent years, to derive the strongest bounds from SN for heavy sterile neutrinos.

A similar remark applies on the relation between SN and cosmological bounds.
It is  reassuring that the bulk of the excluded parameter space overlap. The underlying assumptions in deriving the two classes of bounds are indeed very different. For instance, in non-standard cosmological scenarios with low-reheating temperatures, the BBN bounds can be lifted~\cite{Gelmini:2004ah,Gelmini:2008fq}. Since in astroparticle physics one cannot control experimental conditions, the accumulation of independent ways to probe a certain type of new physics is essential for a broad acceptance of the robustness of the derived bounds.

\vspace{2cm}
\acknowledgements

We warmly thank A. Lella for discussion during the preparation of this work. We are grateful to D.F.G. Fiorillo, M. Ovchynnikov and E. Vitagliano for comments on the manuscript. 
PC and GL thank the Galileo Galilei Institute for Theoretical Physics for 
hospitality during the preparation of part of this work.
This article is based upon work from COST Action COSMIC WISPers CA21106, supported by COST (European Cooperation in Science and Technology).
The work of PC is supported by the European Research Council under Grant No.~742104 and by the Swedish Research Council (VR) under grants 2018-03641 and 2019-02337.
The work of LM is supported by the Italian Istituto Nazionale di Fisica Nucleare (INFN) through the ``QGSKY'' project and by Ministero dell'Universit\`a e Ricerca (MUR).
The work of AM  was partially supported by the research grant number 2022E2J4RK "PANTHEON: Perspectives in Astroparticle and
Neutrino THEory with Old and New messengers" under the program PRIN 2022 funded by the Italian Ministero dell’Universit\`a e della Ricerca (MUR). GL is supported by the European Union’s Horizon 2020 Europe research and innovation programme under the Marie Skłodowska-Curie grant agreement No 860881-HIDDeN.\\
This work is (partially) supported
by ICSC – Centro Nazionale di Ricerca in High Performance Computing,
 Big Data and Quantum Computing, funded by European Union - NextGenerationEU. 
The computational work has been executed on the IT resources of the ReCaS-Bari data center, which have been made available by two projects financed by the MIUR (Italian Ministry for Education, University and Re-search) in the "PON Ricerca e Competitività 2007-2013" Program: ReCaS (Azione I - Interventi di rafforzamento strutturale, PONa3\_00052, Avviso 254/Ric) and PRISMA (Asse II - Sostegno all'innovazione, PON04a2A).

\onecolumngrid
\appendix

\section{Production rates for massive neutrino production via nuclear interactions}
\label{Sec:Implementation of the code}
Here we discuss how to compute the production rates for the neutral current interactions with nucleons ${\nu_\alpha(p_1) + N(p_2)\rightarrow N(p_3)+\nu_4(p_4)}$ and the charged current process with muons $\mu(p_1) + N(p_2)\rightarrow N(p_3)  +\nu_4(p_4)$. 

In general, following the recipe in Ref.~\cite{Hannestad:1995rs}, the nine-dimensional integral for the production rate in Eq.~\eqref{eq:dnde} can be reduced to a three-dimensional integration that can be evaluated numerically. Explicitly,
\begin{equation}
    \frac{{\rm d}^{2}n_4}{{\rm d}E_4 {\rm d}t}= \frac{|U_{\alpha 4}|^2}{(2\pi)^6} p_4\int_0^\infty\frac{p_3^2{\rm d}p_3}{2E_3}\int_0^{p_3+p_4}\frac{p_1^2{\rm d}p_1}{2E_1}\int{\rm d}\cos\theta \,M(p_1,p_3,p_4,\cos\theta)\,f_1\,f_2\,(1-f_3) \,\ ,
    \label{eq:appcoll}
\end{equation}
where $p_i=|\textbf{p}_i|$, $\cos \theta = \textbf{p}_1\cdot \textbf{p}_4/p_1\,p_4$, the integration limits for $\cos\theta$ are expressed in Ref.~\cite{Hannestad:1995rs} and 
\begin{equation}
    M(p_1,p_3,p_4,\cos\theta) = \int dx \,|\mathcal{M}|^2\,,
    \label{eq:dx}
\end{equation}
is an integral that can be analytically evaluated~\cite{Hannestad:1995rs}, with $ x= \textbf{p}_3\cdot \textbf{p}_4/p_3\,p_4$.

We compute the matrix elements for charged current processes $l^-(p_1) + N(p_2)\rightarrow N(p_3)+\nu_4(p_4)$ (see Eqs.~(B1a)-(B1c) in Ref.~\cite{Guo:2020tgx}) without neglecting neither the charged lepton nor the neutrino mass, as usually done for the SM channels. In particular, we define $|\mathcal{M}|^2=\langle |\mathcal{M}|^2\rangle_{VV}+\langle |\mathcal{M}|^2\rangle_{VA}+\langle |\mathcal{M}|^2\rangle_{AA}$, with
\begin{eqnarray}
    \langle |\mathcal{M}|^2\rangle_{VV} &=& 16\, G^2\, G_V^2\, \left[ (p_1\cdot p_2)(p_3\cdot p_4) + (p_2\cdot p_4)(p_1\cdot p_3) - m_2\,m_3\,(p_1\cdot p_4) \right] \,\ , \label{MVV} \\
 \langle |\mathcal{M}|^2\rangle_{VA} &=&  32\,G^2\,G_V\,G_A\,[(p_1\cdot p_2)(p_3\cdot p_4)-(p_2\cdot p_4)(p_1\cdot p_3)] \,\ , \label{MVA} \\
  \langle |\mathcal{M}|^2\rangle_{AA} &=&  16\,G^2\,G_A^2\,[(p_1\cdot p_2)(p_3\cdot p_4)+(p_2\cdot p_4)(p_1\cdot p_3)+m_2\,m_3\,(p_1\cdot p_4)]]  \,\ , \label{MAA} \\
    G&=&G_FV_{ud} \,\ , \\
    G_V&=&\frac{g_V\left(1-\frac{q^2(\gamma_p-\gamma_n)}{4M_N^2}\right)}{\left(1-\frac{q^2}{M_N^2}\right)\left(1-\frac{q^2}{M_V^2}\right)^2} \,\ , \\
    G_A&=&\frac{g_A}{
    \left(1-\frac{q^2}{M_A^2}\right)^2} \,\ . 
\end{eqnarray}
Here, $G_F$ is the Fermi constant, $V_{ud}$ is the up-down entry of the Cabibbo-Kobayashi-Maskawa
matrix, $\gamma_p$ and $\gamma_n$ the magnetic moments of protons and
neutrons, respectively, $g_V=1$ and $g_A=1.27$ the vector and axial vector coupling constant, respectively. In addition, $M_V=840~\mathrm{MeV}$ is the vector mass, $M_A=1~\mathrm{GeV}$ the axial mass, and $M_N$ the nucleon mass, which in the vacuum is $M_N=938~\MeV$, while in the SN plasma is reduced to an effective mass $M_N\sim\mathcal{O}(500)~\MeV$ due to nuclear self-interaction~\cite{Hempel:2014ssa}.
To numerically evaluate Eq.~\eqref{eq:appcoll}, we have defined
\begin{eqnarray}
 p_3\cdot p_4 = && E_3\,E_4 - p_3\,p_4\,x\,, \nonumber \\ 
 p_1\cdot p_4 = && E_1\,E_4 - p_1\,p_4\,\cos\theta\,, \nonumber \\
 p_2\cdot p_4 = && m_4^2 + (E_3\,E_4 - p_3\,p_4\,x) - (E_1\,E_4 - p_1\,p_4,\cos\theta)\,, \nonumber \\
 p_1\cdot p_3 = && (E_3\,E_4 - p_3\,p_4\,x) - (E_1\,E_4-p_1\,p_4\,\cos\theta) + Q/2\,, \nonumber \\
 p_2\cdot p_3 = && (E_1\,E_4-p_1\,p_4\,\cos\theta) + m_3^2 - Q/2\,, \nonumber \\
 p_1\cdot p_2 = && (E_3\,E_4 - p_3\,p_4\,x) - m_1^2 + Q/2\,, \nonumber \\
 Q= &&m_1^2+m_3^2+m_4^2-m_2^2 \,\ .\nonumber
\end{eqnarray}

With the above definition, we can write the three matrix element terms in Eqs.~\eqref{MVV}, \eqref{MVA} and \eqref{MAA} as
\begin{eqnarray}
    \langle |\mathcal{M}|^2\rangle_{VV} = && 8\,G^2\,G_V^2\,[2\,\cos^2\theta\,p_1^2\,p_4^2 + 2\,E_3\,E_4\,(2\,\cos\theta\,p_1\,p_4+m_4^2-m_1^2+Q)- \nonumber \\
    && -E_1\,E_4\,(4\,\cos\theta\,p_1\,p_4+2\,m_4^2+2\,m_2\,m_3+Q)+\cos\theta\,p_1\,p_4(2\,m_4^2+2\,m_2\,m_3+Q)+  \nonumber \\
    && +2\,E_4^2\,(2\,E_3^2-2\,E_1\,E_3+E_1^2)+m_4^2\,Q] -  \\
    && - 16\,G^2\,G_V^2\,p_3\,p_4\,(2\cos\theta p_1\,p_4 + 4\,E_3\,E_4 - 2\,E_1\,E_4 + m_4^2 - m_1^2 + Q)\,x +  32\, G^2\,G_V^2 p_3^2\,p_4^2\,x^2 \nonumber \,\ , \\
    \langle |\mathcal{M}|^2\rangle_{VA} = && 16\,G_A\,G^2\,G_V\,[-2\,E_3\,E_4\,(2\cos\theta\,p_1\,p_4 + m_1^2+m_4^2)+E_1\,E_4\,(4\cos\theta\,p_1\,p_4 + 2\,m_4^2 +Q) - \nonumber \\
 && - (\cos\theta\,p_1\,p_4 + m_4^2)(2\cos\theta\,p_1\,p_4 + Q) + 2\,E_4^2\,E_1 (2E_3 - E_1)] +  \\
 && + 32\,G_A\,G^2\,G_V\,p_3\,p_4\,(2\cos\theta p_1\,p_4 - 2\,E_1\,E_4 + m_1^2 + m_4^2)\,x \nonumber \,\ , \\
 \langle |\mathcal{M}|^2\rangle_{AA} = && 8\,G^2\,G_A^2\,[2\,\cos^2\theta p_1^2\,p_4^2 + 2\,E_3\,E_4 (2\,\cos\theta\,p_1\,p_4 + m_4^2 - m_1^2 +Q) - \nonumber \\
&& - E_1\,E_4 (4\,\cos\theta\,p_1\,p_4 + 2\,m_4^2 - 2\,m_2\,m_3 + Q) + \cos\theta\,p_1\,p_4 (2\,m_4^2 -  2\,m_2\,m_3 + Q) + \nonumber \\
&& + 2\,E_4^2\,(2E_3^2- 2\,E_1\,E_3 + E_1^2) + m_4^2 Q] - \nonumber \\
&& - 16\,G^2\,G_A^2\,p_3\,p_4 (2\cos\theta\,p_1\,p_4 + 4\,E_3\,E_4 - 2\,E_1\,E_4 + m_4^2 - m_1^2 + Q)\,x + 32\, G^2\, G_A^2\,p_3^2\,p_4^2\,x^2\,.
\end{eqnarray}
Finally, to obtain $M(p_1,p_3,p_4,\cos\theta)$ in Eq.~\eqref{eq:appcoll}, we need to analytically integrate $|\mathcal{M}|^2$ over $dx$ as shown in Eq.~\eqref{eq:dx} and discussed in Ref.~\cite{Hannestad:1995rs}. 

The production rate for the neutral-current interaction $\nu_\alpha \,N \to N\, \nu_4\, $ can be computed in a way analogous to the charged-current one, with the replacements~\cite{1985ApJS...58..771B}
\begin{eqnarray}
    G_V&\to& G_V^n=\frac{1}{2} \,\,\,\,\,\, , \,\,\, G_V^p=\frac{1}{2}-2\sin^2\theta_W \,\ , \\
    G_A&\to& G_A^n=\frac{g_A}{2} \,\ , \,\ G_A^p=\frac{g_A}{2} \,\ .
\end{eqnarray}

These general expressions can be used to compute the emissivities for the processes $\nu_4\, N \leftrightarrow \nu_\alpha\, N$ and $\nu_4 \,N \leftrightarrow \mu\, N$. On the other hand, details on the computation of the production rates for the other processes shown in Tab.~\ref{tab:scatteringt} can be found in Ref.~\cite{Mastrototaro:2019vug}.

\bibliographystyle{bibi}
\bibliography{biblio.bib}

\end{document}